\documentclass[11pt]{article}
\usepackage[a4paper,margin=29mm]{geometry}
\usepackage[T1]{fontenc}
\usepackage{lmodern}
\usepackage{amsmath,amssymb,amsthm,mathtools}
\usepackage{microtype}
\usepackage{tikz}
\usetikzlibrary{arrows.meta}
\usepackage[hidelinks]{hyperref}
\newcommand{\PaperAuthor}{Obinna Okechukwu}
\hypersetup{
  pdftitle={Sharper Bounds for Graphical Two-Choice Allocation},
  pdfauthor={\PaperAuthor},
  pdfsubject={Joint exponential moments, graphical allocation, and stochastic cut workloads},
  pdfkeywords={graphical allocation, load balancing, orthogonal contrasts, stochastic cut workloads}}
\newtheorem{theorem}{Theorem}[section]
\newtheorem{lemma}[theorem]{Lemma}
\newtheorem{proposition}[theorem]{Proposition}
\newtheorem{corollary}[theorem]{Corollary}
\theoremstyle{remark}
\newtheorem{remark}[theorem]{Remark}
\newcommand{\E}{\mathbb E}
\newcommand{\OPT}{\operatorname{OPT}}
\newcommand{\OFF}{\operatorname{OFF}}
\newcommand{\W}{\mathcal W}
\newcommand{\1}{\mathbf 1}
\DeclareMathOperator{\sgn}{sgn}
\DeclareMathOperator{\Gap}{Gap}
\newcommand{\pos}[1]{[#1]_+}
\newcommand{\dd}{\overline d}
\newcommand{\cmin}{c_{\min}}
\allowdisplaybreaks[1]
\title{Sharper Bounds for Graphical Two-Choice Allocation}
\author{\PaperAuthor}
\date{}

\begin{document}
\maketitle
\begin{abstract}
In graphical two-choice allocation, vertices are bins, and each arriving ball
must be placed irrevocably at one endpoint of a uniformly random edge. The
objective is to keep the maximum-minus-minimum load gap small at every time. On
cycles, endpoint-local monotone rules incur polynomial gaps at long time
horizons, whereas global rules can achieve polylogarithmic gaps. We prove a
joint exponential-moment inequality for the orthogonal contrasts of a binary
partition tree, controlling all contrasts along a root-to-leaf path
simultaneously. Applied to the Bansal--Feldheim strategy, it improves the gap
on every $\kappa$-edge-connected $d$-regular graph from $O((d/\kappa)\log^4
n\log\log n)$ to $O((d/\kappa)\log^3 n\log\log n)$. We give an explicit
strategy on the $n$-cycle with expected gap $O(\log^2 n)$ uniformly over
deterministic times and exponential tails at each such time. An Euler-tour
projection yields $O(\bar d\log^2(2M))$ for every connected graph with $M$
edges and degree-proportional capacities. For lower bounds, we introduce a
stochastic cut workload: the expected largest normalized overload forced by the
requests, plus an indivisibility term. For every online integral strategy, the supremum over time of
the expected gap is at least one sixth of this benchmark. Regular rings of cliques
attain $\Omega((d/\kappa)\log n)$ along an explicit family, improving the prior
additive lower bound. With unequal capacities, the workload need not control
the gap: one family has constant relative cut slack, bounded workload and gap
$\Omega(\log\log n)$; another has workload at most $20$ but forces gap
$\Omega(\log n)$ even for clairvoyant fractional allocations that remain
irrevocable.
\end{abstract}



\section{Introduction}

Let $G=(V,E)$ be a finite graph. At each step, a uniformly random edge is
offered and one new ball must be placed on one of its endpoints. The decision
may use the entire allocation history, but the ball cannot subsequently be
moved. The model describes load balancing in which each job may be served
by only one of two related resources, such as adjacent servers in a
network. We study the gap between the largest and smallest loads, uniformly
in the number of arrivals. The greedy rule, which places each ball on the
less loaded offered endpoint, has logarithmic gap on expanders~\cite{PTW15}.
The question is how much a rule using the entire load vector can improve on
local allocation for poorly expanding graphs such as cycles.

The classical power-of-two-choices theorem of Azar, Broder, Karlin and
Upfal~\cite{ABKU99}, and its heavily loaded extension by Berenbrink, Czumaj,
Steger and V\"ocking~\cite{BCSV06}, concern the largest load above the average.
The maximum-minus-minimum gap is a different quantity: even on the complete
graph, its natural scale is logarithmic rather than doubly logarithmic, since
some vertex receives no offer for order $n\log n$ steps; Bansal and
Feldheim~\cite{BF22} record this $\Omega(\log n)$ bound for every graph.
Kenthapadi and Panigrahy~\cite{KP06} initiated the study of two-choice
allocation on graphs. Peres, Talwar and Wieder~\cite{PTW15} bounded the gap
of greedy allocation by the edge expansion, giving $O(n\log n)$ on the
cycle. Olesker-Taylor, Sauerwald and Zanetti~\cite{OSZ26} obtain
an $O(n)$ expected cycle gap in equilibrium for a synchronous centered-load
process. Their centering subtracts $1/n$ from every load after each step,
which preserves all pairwise differences; the equilibrium estimate is not
a uniform-in-time bound from the empty state. Their principal model is a
continuous-time queueing process with removals.
Polynomial growth of the greedy gap on the cycle was conjectured
in~\cite{BF22}. Okechukwu~\cite{Oke26} subsequently proved that
every endpoint-local monotone rule, including greedy and allowing
edge-dependent thresholds and fresh randomization, has expected gap
$\Omega(\min\{\sqrt n,(k/n)^{1/4}\})$ after $k\ge1$ allocations.
In particular, the lower bound is $\Omega(\sqrt n)$ for $k\ge n^3$.
The result does not apply to strategies that use the whole load vector.
For the maximum in-degree objective under i.i.d.\ uniform edges of a base
graph, Bansal, Prabhu, Singla and Sundaram~\cite{BPSS26} give an
$O(\log\log n)$-competitive online algorithm for every base graph, tight
already for the complete graph; that multiplicative objective differs from
the additive max-minus-min gap studied here.

Bansal and Feldheim~\cite{BF22} gave a global strategy on every connected
$d$-regular graph of edge connectivity $\kappa$. Their bound is
\[
 O\!\left(\alpha_G\frac d\kappa\log^2 n\right),
\]
where $\alpha_G$ is the congestion parameter of the cut-based routing
decomposition used in their construction. The decomposition theorem of
Harrelson, Hildrum and Rao~\cite{HHR03}, building on R\"acke~\cite{Racke02},
provides $\alpha_G=O(\log^2 n\log\log n)$. For cycles, Bansal and Feldheim
record $\alpha_G=O(\log n)$, so their analysis implies a cycle bound of
$O(\log^3 n)$.

We improve the analysis of the same general-graph strategy to
\begin{equation}\label{eq:intro-general}
 O\!\left(\alpha_G\frac d\kappa\log n\right)
 =O\!\left(\frac d\kappa\log^3 n\log\log n\right).
\end{equation}
The precise estimate, including its fixed-time exponential tail, is
Theorem~\ref{thm:regular}. We also give an explicit cycle strategy with
$O(\log^2 n)$ gap in Theorem~\ref{thm:cycle}. Its implementation maintains a
binary interval tree and uses $O(\log n)$ arithmetic operations per arrival.
Proposition~\ref{prop:euler} projects this strategy through an Euler tour of
the doubled graph. With degree-proportional capacities, the resulting bound on
every connected graph is $O(\dd\log^2(2M))$, where $\dd=2M/n$.

The improvement in~\eqref{eq:intro-general} comes from controlling all
contrasts on a root-to-leaf chain together. Bansal and Feldheim realize
inward drift in orthogonal sibling contrasts, bound those contrasts
separately, and sum along a chain. We instead prove
\[
 \E\prod_{I\in J}\cosh(\mu\delta_I(t))\le 2^{|J|}
\]
for every set $J$ of hierarchy nodes. Auxiliary signs express the product as
one exponential. Orthogonality then bounds its quadratic increment, even
though the contrasts are dependent and the allocation rule may depend on
the entire history. A single maximum over the leaves replaces separate
contrast thresholds followed by a depth sum. Exponential and
hyperbolic-cosine potentials are established tools in balanced
allocation~\cite{PTW15,LS24}, and exponential potentials over many
coordinates also drive online vector balancing~\cite{ALS21}; the estimate
here controls their product in orthogonal hierarchy coordinates.

On the lower-bound side, a cut records the work that the requests force into
one set of vertices. We take the expected largest such normalized overload
over all cuts and all prefixes, and add the indivisibility scale
$1/\cmin$. The resulting stochastic cut workload $\W$ is defined
in~\eqref{eq:workload}. Theorem~\ref{thm:workload} proves
$\OPT\ge\W/6$, where $\OPT$ is the least value of $\sup_t\E H_t$ achievable by an
online strategy.
The supremum in $\W$ is inside the expectation. A reverse first-crossing
argument connects it to the supremum of deterministic-time expectations in
$\OPT$.

The workload also exposes the effect of many disjoint bottlenecks.
Theorem~\ref{thm:rings} constructs regular rings of cliques for which
\[
 \OPT\ge \frac{d}{128\kappa}\log N,
\]
where $N$ is the number of cliques. Taking $N$ equal to the clique size gives
$\Omega((d/\kappa)\log n)$, compared with the separate
$\Omega(d/\kappa+\log n)$ lower bounds of Bansal and
Feldheim~\cite{BF22}. The proof uses pairwise negative correlation of
disjoint-cut hitting events; it does not assume independent bottlenecks.
The companion workload upper bound uses classical random-contraction
cut counting~\cite{Karger00}.

For general i.i.d.\ edge laws and unequal capacities, bounded cut workload does not suffice.
Theorem~\ref{thm:slack} gives a family with relative cut slack $1/2$ and
$\W\le10$, but online integral gap $\Omega(\log\log n)$.
Theorem~\ref{thm:hub} gives $\W\le20$ and gap $\Omega(\log n)$ even for
fractional assignments chosen with knowledge of the whole request
sequence, provided assignments remain irrevocable. The proof bounds the
total mass available to serve vertices missed by an internally forced
request stream. Heterogeneous sampling was studied by
Wieder~\cite{Wieder07}; Los, Sauerwald and Sylvester~\cite{LSS23} study a
model with memory that supplies an additional allocation choice.
Here the eligible set remains the current request, and the capacities are
specified separately from the request probabilities.

Our upper bounds hold at each deterministic time, with constants and a
single strategy independent of that time. They do not bound the largest
gap over an infinite trajectory. The heterogeneous counterexamples have
neither equal nor offer-proportional capacities. Workload sufficiency for those two
classes remains open, as does an $O(\log n)$ bound on the cycle. Models
that average endpoint loads after an arrival, such as
\cite{ANS20}, permit a different evolution of existing loads.

\section{Model and hierarchical contrasts}\label{sec:model}

\subsection{Requests, capacities, and time conventions}

Throughout, $V$ is finite and $n=|V|\ge2$. The conventions in this
section apply to all subsequent statements unless explicitly changed.
Time indices are integers; thresholds such as $x$ and $z$ may be real.
A request is a nonempty subset
of $V$ of size at most two. The requests $e_1,e_2,\ldots$ are independent
with a common law $p$. An online integral strategy chooses one member of
$e_t$ using the requests and decisions through time $t$ and private
randomness independent of future requests; that is, it is
non-anticipating. The load vector starts at
$L_0=0$. Let $\mathcal F_t$ contain the requests, decisions, and private
randomness used through time $t$. In the graph results, the graph is simple and connected and
$p$ is uniform on its edges.

Let $c_v>0$ be the capacity of vertex $v$. Write
$c(S)=\sum_{v\in S}c_v$, $C=c(V)$, and $\cmin=\min_v c_v$. Define
\begin{equation}\label{eq:objective}
 H_c(L)=\max_{v\in V}\frac{L_v}{c_v}
       -\min_{v\in V}\frac{L_v}{c_v},
 \qquad
 \OPT(G,p,c)=\inf_{\mathcal A}\sup_{t\ge0}\E H_c(L_t^{\mathcal A}).
\end{equation}
The infimum is over online integral strategies. We abbreviate
$H_t=H_c(L_t)$. For equal capacities, $c_v=1$ and we also write
$\Gap(L)=H_c(L)$. For uniform edge requests we use the degree-proportional capacities
\[
 c_v=\frac{\deg_G(v)}{\dd},\qquad \dd=\frac{2M}{n},\qquad M=|E|,
\]
so $C=n$. More generally, for a law supported on pairs the offer-proportional
capacities are $c_v=(n/2)\Pr(v\in e_1)$, when these are positive; for
uniform edge requests they are the degree-proportional capacities.

The capacity-weighted mean normalized load at time $t$ is $t/C$.
Consequently the centered coordinates
$X_v(t)=L_t(v)/c_v-t/C$ satisfy
\begin{equation}\label{eq:centered-range}
 |X_v(t)|\le H_t \qquad(v\in V).
\end{equation}
A fractional trajectory may split a request's unit mass between its
eligible vertices. It is \emph{irrevocable} if all assigned fractions stay
at their destinations. We state explicitly when such trajectories, or
trajectories allowed to depend on future requests, are considered.
An instance is \emph{strictly feasible} if
$\Pr(e_1\subseteq S)<c(S)/C$ for every proper nonempty $S$. The
\emph{slack} of $S$ is $c(S)/C-\Pr(e_1\subseteq S)$, and its
\emph{relative slack} is the slack divided by
$\min\{c(S),c(S^c)\}/C$. The \emph{relative cut slack} of the
instance is the minimum relative slack over proper nonempty $S$.
All logarithms are natural unless a base is displayed.

\subsection{Averages on a binary partition tree}

A hierarchy records average loads at successively finer scales. Differences
between sibling averages are suited to this purpose because their
coefficient vectors are orthogonal even when the children have unequal
sizes.

Let $\mathcal T$ be a full rooted binary partition tree with root $V$ and
singleton leaves. Identify each node with its vertex set. For an internal
node $I$ with children $I_0,I_1$, put
\begin{equation}\label{eq:contrasts}
 h_I=\frac{\1_{I_0}}{|I_0|}-\frac{\1_{I_1}}{|I_1|},
 \qquad \delta_I(t)=\langle h_I,L_t\rangle,
 \qquad \nu_I=\|h_I\|_2^2=\frac1{|I_0|}+\frac1{|I_1|}.
\end{equation}
Let $\mathcal C_v$ be the internal nodes on the root-to-$v$ path, and define
\begin{equation}\label{eq:tree-parameters}
 D=\max_v|\mathcal C_v|,\qquad
 B_*=\max_v\sum_I|h_I(v)|.
\end{equation}
Up to normalization, the vectors $h_I$ form the unbalanced Haar basis
of the partition tree~\cite{GS97}.
The quantity $D$ counts the contrasts on a root-to-leaf path. The
quantity $B_*$ bounds the total change in their signed sum when a single
ball is allocated.

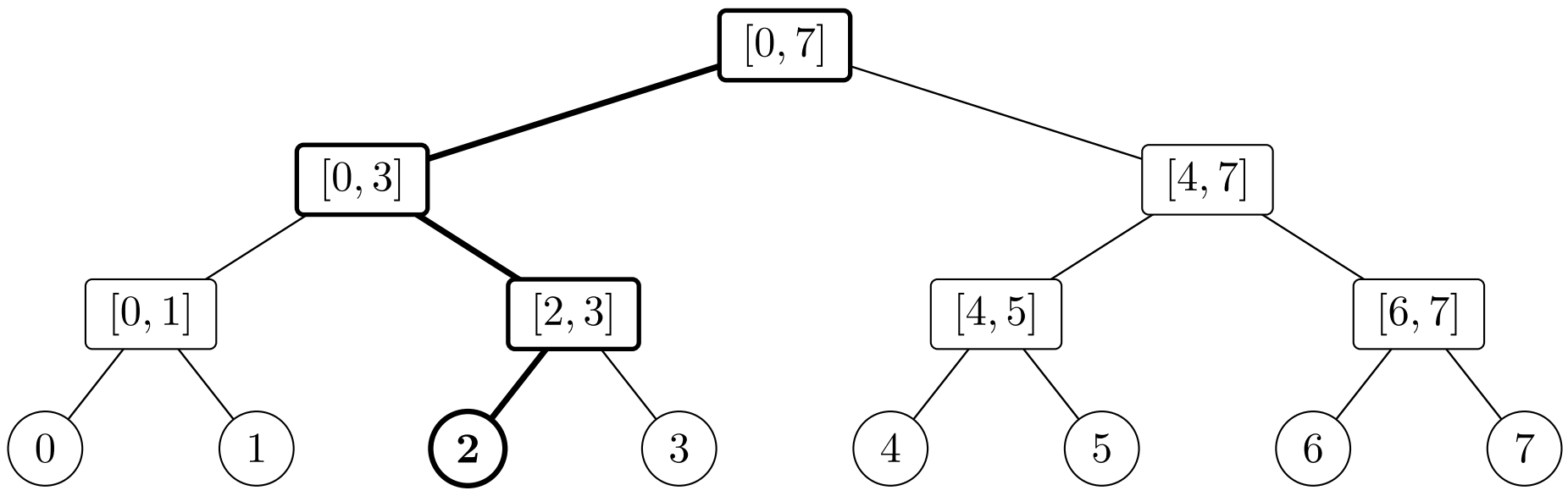
\begin{figure}[htbp]
\centering
\begin{tikzpicture}[
  x=1.65cm,y=1.05cm,
  every node/.style={font=\fontsize{9.5}{11}\selectfont},
  cluster/.style={draw,rounded corners=1.5pt,fill=white,
    inner xsep=5pt,inner ysep=3pt},
  leaf/.style={circle,draw,fill=white,minimum size=5.8mm,inner sep=0pt},
  branch/.style={line width=.45pt},
  selected/.style={line width=1.35pt}
]
  \coordinate (root) at (3.5,3);
  \coordinate (left) at (1.5,2);
  \coordinate (right) at (5.5,2);
  \foreach \i/\x in {0/.5,1/2.5,2/4.5,3/6.5}
    \coordinate (pair\i) at (\x,1);
  \foreach \i in {0,...,7}
    \coordinate (v\i) at (\i,0);
  \draw[branch] (root)--(left) (root)--(right);
  \draw[branch] (left)--(pair0) (left)--(pair1)
                (right)--(pair2) (right)--(pair3);
  \draw[branch] (pair0)--(v0) (pair0)--(v1)
                (pair1)--(v2) (pair1)--(v3)
                (pair2)--(v4) (pair2)--(v5)
                (pair3)--(v6) (pair3)--(v7);
  \draw[selected] (root)--(left)--(pair1)--(v2);
  \node[cluster,line width=1pt] at (root) {$[0,7]$};
  \node[cluster,line width=1pt] at (left) {$[0,3]$};
  \node[cluster] at (right) {$[4,7]$};
  \node[cluster] at (pair0) {$[0,1]$};
  \node[cluster,line width=1pt] at (pair1) {$[2,3]$};
  \node[cluster] at (pair2) {$[4,5]$};
  \node[cluster] at (pair3) {$[6,7]$};
  \foreach \i in {0,1,3,4,5,6,7}
    \node[leaf] at (v\i) {$\i$};
  \node[leaf,line width=1.2pt] at (v2) {$\mathbf2$};
\end{tikzpicture}
\caption{A dyadic hierarchy on eight vertices. The bold chain contains
exactly the contrasts needed to reconstruct the deviation of vertex~\(2\)
from the global average. The proof controls their sum through one joint
moment, before taking a maximum over the leaves.}
\label{fig:hierarchy}
\end{figure}

\begin{lemma}\label{lem:haar}
The vectors $h_I$ are mutually orthogonal and have zero coordinate sum.
For every load vector $L$,
\begin{equation}\label{eq:reconstruction}
 L_v-\frac{L(V)}n
   =\sum_{I\in\mathcal C_v}\frac{h_I(v)}{\nu_I}\,
                  \langle h_I,L\rangle .
\end{equation}
In particular, if $A_v=\sum_{I\in\mathcal C_v}|\langle h_I,L\rangle|$, then
$\Gap(L)\le2\max_v A_v$. If every split has equal child sizes, then
$\Gap(L)\le\max_v A_v$.
\end{lemma}

\begin{proof}
We compare averages across one parent-child edge and then telescope to a
leaf. Distinct hierarchy sets are disjoint or nested. In the nested case
the coarser contrast is constant on the support of the finer one, whose
coordinate sum is zero. This proves orthogonality.

Write $a=|I_0|$, $b=|I_1|$ and let $\overline L(I)=L(I)/|I|$.
The two parent-child differences are
\[
 \overline L(I_0)-\overline L(I)=\frac b{a+b}\delta_I,\qquad
 \overline L(I_1)-\overline L(I)=-\frac a{a+b}\delta_I.
\]
These coefficients equal $h_I(v)/\nu_I$ on the corresponding child.
Summing along the path proves~\eqref{eq:reconstruction}.
Their absolute values are at most one, so every centered load has absolute
value at most $A_v$. Taking the range gives the first gap bound.
For equal splits the absolute coefficients are $1/2$, giving the second.
\end{proof}

\section{Joint exponential moments}\label{sec:moments}

Separate tail bounds for the contrasts give a threshold at every level.
The following estimate controls their sum before maximizing over leaves.
Its hypothesis is a drift condition in the original time scale; no
independence or autonomous evolution of the contrasts is assumed.

\begin{theorem}\label{thm:joint}
Let $\mathcal T$ be a binary hierarchy on $n$ vertices with $B_*\le8$.
Consider an adapted process with $L_0=0$ and
$L_{t+1}-L_t=\mathbf e_{V_{t+1}}$ at each step, where $\mathbf e_v$ is the
unit vector at $v$. Suppose the conditional probabilities
$q_t(v)=\Pr(V_{t+1}=v\mid\mathcal F_t)$ satisfy $q_t(v)\le2/n$
for every $t\ge0$ and $v$, and for some $0<\eta\le1/4$,
\begin{equation}\label{eq:inward}
 \sgn(\delta_I(t))\,\E[\delta_I(t+1)-\delta_I(t)\mid\mathcal F_t]
       \le-\frac{\eta\nu_I}{n}
       \quad\text{whenever }\delta_I(t)\ne0.
\end{equation}
Set $\mu=\eta/4$. For every set $J$ of internal nodes and every
deterministic time $t$,
\begin{equation}\label{eq:joint-moments}
 \E\prod_{I\in J}\cosh(\mu\delta_I(t))\le2^{|J|},
 \qquad
 \E\exp\!\left(\mu\sum_{I\in J}|\delta_I(t)|\right)\le4^{|J|}.
\end{equation}
In particular, with $A=\log n+D\log4$,
\begin{align}
 \sup_{t\ge0}\E\Gap(L_t)&\le\frac8\eta A,
       \label{eq:joint-mean}\\
 \Pr\!\left\{\Gap(L_t)>\frac8\eta(A+z)\right\}&\le e^{-z}
       \quad(z\ge0). \label{eq:joint-tail}
\end{align}
For hierarchies with equal splits, the coefficient $8$ in both bounds
can be replaced by $4$.
\end{theorem}

\begin{proof}
We express the product of coshes as an average of exponentials so that
orthogonality can control the whole one-step quadratic term.
Put $P_J=\prod_{I\in J}\cosh(\mu\delta_I)$ and let
$(\sigma_I)_{I\in J}$ be independent uniform signs, used only for this
algebraic averaging. Then
\[
 P_J=\E_\sigma\exp\!\left(\mu\sum_{I\in J}\sigma_I\delta_I\right).
\]
If the next ball goes to $v$, its contrast increment is
$\Delta\delta_I=h_I(v)$. Hence
$|\mu\sum_{I\in J}\sigma_I\Delta\delta_I|\le\mu B_*\le1/2$.
For every fixed sign vector, Lemma~\ref{lem:haar} and $q_t(v)\le2/n$ give
\begin{align}
 \E\!\left[\left(\sum_{I\in J}\sigma_I\Delta\delta_I\right)^2
                    \,\middle|\,\mathcal F_t\right]
 &\le\frac2n\sum_v\left(\sum_{I\in J}\sigma_I h_I(v)\right)^2
 \nonumber\\
 &=\frac2n\sum_{I\in J}\nu_I. \label{eq:quadratic}
\end{align}
This bound is uniform in the auxiliary signs. It therefore remains valid
when multiplied by their current positive exponential weights and averaged.

For $|x|\le1/2$, the integral remainder for the exponential gives
$e^x\le1+x+x^2$. Applying it to the one-step exponent increment yields
\begin{align}
 \E[\Delta P_J\mid\mathcal F_t]
 &\le
 \mu\sum_{I\in J}\sinh(\mu\delta_I)P_{J\setminus\{I\}}\,
                  \E[\Delta\delta_I\mid\mathcal F_t]
 \nonumber\\
 &\hspace{7mm}
 +\frac{2\mu^2}{n}\left(\sum_{I\in J}\nu_I\right)P_J .
 \label{eq:product-drift}
\end{align}
The term containing a zero contrast vanishes, so
\eqref{eq:inward} applies to every nonzero contribution.
Since $|\sinh x|\ge\cosh x-1$ and $2\mu^2=\mu\eta/2$, define
\[
 r_I=\frac{\mu\eta\nu_I}{2n},\qquad R_J=\sum_{I\in J}r_I
\]
to obtain the recursion
\begin{equation}\label{eq:subset-recursion}
 \E[P_J(t+1)\mid\mathcal F_t]
 \le(1-R_J)P_J(t)
       +2\sum_{I\in J}r_I P_{J\setminus\{I\}}(t).
\end{equation}
There are $n-1$ internal nodes and $\nu_I\le2$, whence
$0\le R_J<\mu\eta\le1/64$. In particular the first coefficient is
nonnegative.

Induct on $|J|$, with the induction hypothesis asserted at every time.
The empty product is one. For a nonempty $J$, the smaller-subset bounds
in~\eqref{eq:subset-recursion} imply
\[
 \E P_J(t+1)\le(1-R_J)\E P_J(t)+R_J2^{|J|}.
\]
The initial value is $P_J(0)=1$. Induction on $t$ now proves the first
inequality in~\eqref{eq:joint-moments}. The inequality
$e^{|x|}\le2\cosh x$ proves the second.
All moments at a finite time exist because only finitely many load
vectors are reachable by then.

For each leaf set $A_v(t)=\sum_{I\in\mathcal C_v}|\delta_I(t)|$.
The second bound in~\eqref{eq:joint-moments} gives
\[
 \E e^{\mu\max_v A_v(t)}
   \le\sum_v\E e^{\mu A_v(t)}\le n4^D=e^A.
\]
Jensen's inequality bounds $\E\max_v A_v(t)$ by $A/\mu$.
Markov's inequality gives
$\Pr\{\max_v A_v(t)>(A+z)/\mu\}\le e^{-z}$.
Apply the two reconstruction bounds in Lemma~\ref{lem:haar}.
\end{proof}

\section{Allocation strategies on cycles and general graphs}\label{sec:upper}

\subsection{An explicit cycle rule}

On a cycle, the cumulative sum of a contrast is a feasible flow with a
simple shape. Allocating according to a signed sum of these flows produces
the prescribed inward drifts simultaneously.

Label $C_n$ by $0,\ldots,n-1$ in cyclic order, and split the interval
$[0,n-1]$ recursively into consecutive halves whose sizes differ by at most
one, taking $I_0$ to be the left child and $I_1$ the right child.
Its depth is $D=\lceil\log_2 n\rceil$. For an internal interval $I$,
define the tent on the edge $e_j=\{j,j+1\bmod n\}$ by
\begin{equation}\label{eq:tent}
 T_I(j)=\sum_{v=0}^{j}h_I(v),\qquad 0\le j<n.
\end{equation}
The tent increases from zero to one across the first child and decreases
to zero across the second. It vanishes outside the internal edges of $I$,
including the wrapping edge $e_{n-1}$. At each depth, at most one tent
is nonzero on a given edge.

\begin{figure}[tb]
\centering
\begin{tikzpicture}[x=1.45cm,y=1.0cm,font=\small]
  \draw[->] (-.35,0)--(7.45,0) node[right]{$j$};
  \draw[->] (-.35,0)--(-.35,1.45);
  \draw[gray!55,densely dotted] (-.35,1)--(7.15,1);
  \node[left] at (-.35,1){$1$};
  \draw[very thick] (0,.25)--(1,.5)--(2,.75)--(3,1)--(4,.75)--(5,.5)--(6,.25)--(7,0);
  \foreach \x/\y in {0/.25,1/.5,2/.75,3/1,4/.75,5/.5,6/.25,7/0}
    \fill (\x,\y) circle (1.4pt);
  \node[anchor=west] at (4.6,1.28){$T_{[0,7]}$};
  \draw[thick,dashed] (0,.5)--(1,1)--(2,.5)--(3,0)--(7,0);
  \node[anchor=west] at (0,1.28){$T_{[0,3]}$};
  \foreach \j in {0,...,7} \node[below] at (\j,-.03){$\j$};
\end{tikzpicture}
\caption{Two exact tent profiles on $C_8$. Their discrete differences are
the root and left-child contrast vectors. The value at edge $7$, which
joins vertices $7$ and $0$, is zero. At most one tent per level uses an
edge, so dividing their signed sum by the depth gives a valid bias.}
\label{fig:tents}
\end{figure}
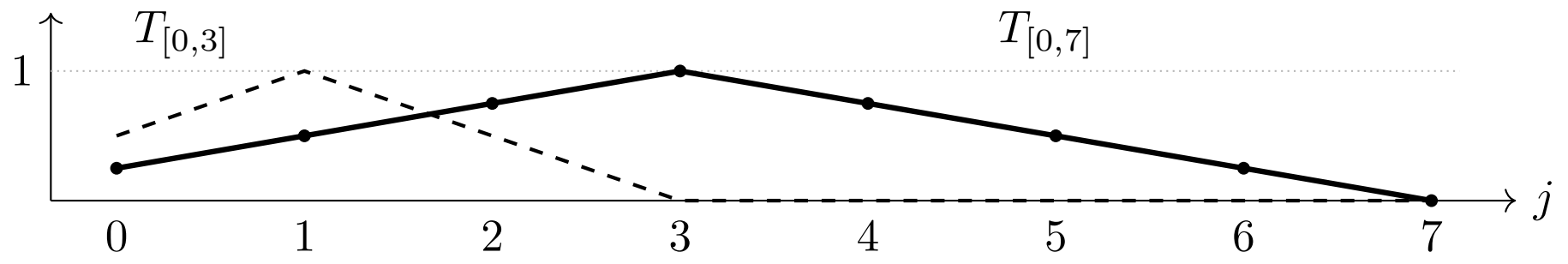

\begin{theorem}\label{thm:cycle}
For $n\ge3$, consider independent, uniformly distributed edge requests
on $C_n$ with unit capacities, starting from $L_0=0$, and put
$D=\lceil\log_2 n\rceil$.
On request $e_j$, assign the
ball to $j$ with probability $(1+b_j)/2$, where
\begin{equation}\label{eq:cycle-rule}
 b_j=-\frac1D\sum_I\sgn(\delta_I)T_I(j),\qquad \sgn(0)=0.
\end{equation}
This online rule is valid and satisfies
\begin{align}
 \sup_{t\ge0}\E\Gap(L_t)
 &\le16D(\log n+D\log4), \label{eq:cycle-mean}\\
 \Pr\{\Gap(L_t)>16D(\log n+D\log4+z)\}
 &\le e^{-z}\quad(z\ge0). \label{eq:cycle-tail}
\end{align}
The probability bound holds for every deterministic $t\ge0$.
If $n=2^D$, both occurrences of $16$ can be replaced by $8$.
Each arrival uses $O(\log n)$ arithmetic operations.
\end{theorem}

\begin{proof}
We check edge feasibility first, since a formal vertex drift alone would
not specify a valid allocation rule. Each tent lies in $[0,1]$ and at most
$D$ tents meet an edge, so $|b_j|\le1$.
The two edges incident to vertex $v$ give its conditional probability
\begin{align}
 q_t(v)
 &=\frac1n+\frac{b_v-b_{v-1}}{2n}
 \nonumber\\
 &=\frac1n-\frac1{2Dn}\sum_I\sgn(\delta_I)h_I(v).
 \label{eq:cycle-marginal}
\end{align}
Indices are cyclic; $T_I(n-1)=0$ makes the difference identity valid also
at $v=0$. A vertex can be selected only when an incident edge arrives,
so $q_t(v)\le2/n$. Orthogonality gives
\[
 \E[\Delta\delta_I\mid\mathcal F_t]
 =-\frac{\nu_I}{2Dn}\sgn(\delta_I).
\]

Along an ancestor chain, a child has at most $\lceil s/2\rceil$ of a
parent's $s\ge2$ vertices. The size ratio is at most $2/3$, apart from
the sharper ratio $1/2$ when $s=2$. Reading the chain upwards therefore gives
$B_*\le\sum_{j\ge0}(2/3)^j=3$.
Apply Theorem~\ref{thm:joint} with $\eta=1/(2D)\le1/4$.
Equal splits give its stronger reconstruction constant.

Maintain the aggregate load at every hierarchy node. Evaluating the bias
uses at most one tent per depth, and placing the ball changes only the
aggregates on its leaf chain. Both operations use $O(D)$ arithmetic
operations. This claim does not bound the bit length of the loads.
\end{proof}

\begin{remark}
The bounds above concern asymptotic guarantees. They do not compare the
finite-size performance of this rule with greedy allocation. The rule is
not of the endpoint-local form covered by~\cite{Oke26}: its bias on an edge
depends on aggregate loads of the hierarchy intervals containing that
edge.
\end{remark}

\subsection{The Bansal--Feldheim strategy}

The hierarchy used for general graphs need not have equal child sizes.
Its geometric contraction is enough to bound $B_*$ independently of its
depth.

\begin{lemma}\label{lem:shrinking}
Suppose that, whenever node $A$ is an ancestor of node $B$ at distance
$s$ in a full binary partition tree on $n$ vertices,
\begin{equation}\label{eq:two-level}
 |B|\le(3/4)^{\lfloor s/2\rfloor}|A|.
\end{equation}
Then
\[
 B_*\le8,\qquad
 D\le2\lfloor\log_{4/3}n\rfloor+1.
\]
\end{lemma}

\begin{proof}
We sum reciprocals of ancestor sizes because those are exactly the
coefficients changed by a ball at a leaf. Write the ancestors of the
leaf containing $v$ as
$A_0,A_1,\ldots,A_\ell=V$. For $j\ge1$,
\[
 |h_{A_j}(v)|=\frac1{|A_{j-1}|}
        \le(3/4)^{\lfloor(j-1)/2\rfloor}.
\]
Their sum is at most $2\sum_{r\ge0}(3/4)^r=8$.
Applying~\eqref{eq:two-level} from $V$ to the leaf $A_0$ gives
$1\le |A_0|\le n(3/4)^{\lfloor\ell/2\rfloor}$,
which bounds the depth.
\end{proof}

\begin{figure}[htbp]
\centering
\begin{tikzpicture}[
  x=1cm,y=1cm,
  every node/.style={font=\fontsize{9.5}{11}\selectfont},
  mass/.style={circle,draw,fill=white,minimum size=7mm,inner sep=0pt},
  inserted/.style={mass,dashed},
  branch/.style={line width=.5pt}
]
  \coordinate (oroot) at (-3.9,2.4);
  \coordinate (o7) at (-5.8,0);
  \coordinate (o4) at (-4.55,0);
  \coordinate (o3) at (-3.25,0);
  \coordinate (o2) at (-2,0);
  \draw[branch] (oroot)--(o7) (oroot)--(o4)
                (oroot)--(o3) (oroot)--(o2);
  \node[mass] at (oroot) {$16$};
  \node[mass] at (o7) {$7$};
  \node[mass] at (o4) {$4$};
  \node[mass] at (o3) {$3$};
  \node[mass] at (o2) {$2$};
  \node at (-3.9,-.65) {Original child clusters};
  \coordinate (rroot) at (3.2,2.4);
  \coordinate (r7) at (1.8,1.6);
  \coordinate (r9) at (4.3,1.6);
  \coordinate (r4) at (3.25,.8);
  \coordinate (r5) at (5.1,.8);
  \coordinate (r3) at (4.45,0);
  \coordinate (r2) at (5.75,0);
  \draw[branch] (rroot)--(r7) (rroot)--(r9)
                (r9)--(r4) (r9)--(r5)
                (r5)--(r3) (r5)--(r2);
  \node[mass] at (rroot) {$16$};
  \node[mass] at (r7) {$7$};
  \node[inserted] at (r9) {$9$};
  \node[mass] at (r4) {$4$};
  \node[inserted] at (r5) {$5$};
  \node[mass] at (r3) {$3$};
  \node[mass] at (r2) {$2$};
  \node at (3.8,-.65) {Binary refinement};
  \draw[-{Latex[length=2mm]},line width=.55pt]
    (-.9,1.2)--(.65,1.2);
\end{tikzpicture}
\caption{The binary refinement used in the general-graph argument.
Numbers are cluster sizes; dashed circles are inserted clusters.
The original children need not have equal sizes. The refinement preserves
geometric contraction over every two levels, which bounds the sum of
contrast coefficients along a leaf by a constant.}
\label{fig:refinement}
\end{figure}
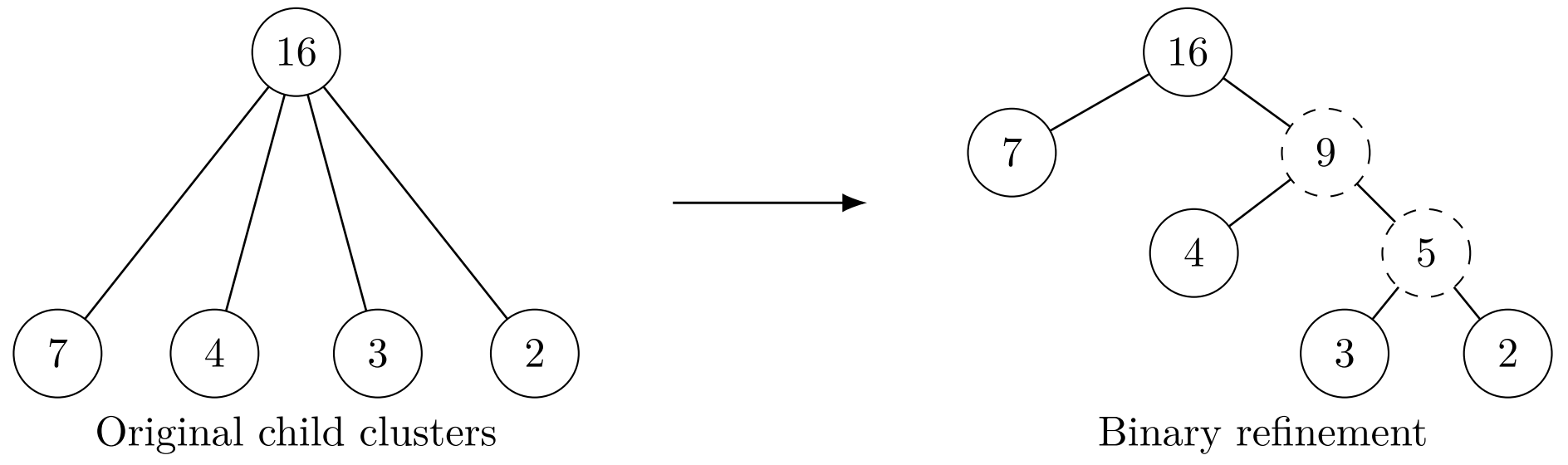

We use the following precise output of the construction
in~\cite{BF22}. All section, equation, and lemma numbers in this paragraph
refer to its full version, arXiv:2106.06051v2.
Section~2.2.2 constructs a binary refinement of the routing tree, and
Claim~3.1 proves~\eqref{eq:two-level}.
For $\beta=1/(8\alpha_G)$, the routed antisymmetric flows
$g_I(u,v)=-g_I(v,u)$ satisfy
\begin{equation}\label{eq:bf-flows}
 \sum_I|g_I(u,v)|\le1,\qquad
 \sum_{u:\,uv\in E}g_I(u,v)=-\beta\kappa h_I(v).
\end{equation}
The divergence is specified in its equation~(2), and Lemma~3.4 proves
the edge budget after routing. These are properties of actual edge
flows, not only desired vertex marginals.
The online rule biases an offered edge toward $v$ by
\[
 b_t(u,v)=\sum_I g_I(u,v)\sgn(\delta_I(t)).
\]
Thus it selects $v$ with probability $(1+b_t(u,v))/2$.
This is equation~(8) of the cited version, with $\sgn(0)=0$.

\begin{theorem}\label{thm:regular}
Let $G$ be a connected simple $d$-regular graph on $n\ge2$ vertices,
with edge connectivity $\kappa$. Use unit capacities and let
edges be offered independently and uniformly. For the Bansal--Feldheim
decomposition and strategy
just specified, let $\alpha_G\ge1$ be its congestion parameter and $D$ its
binary depth. Starting empty, this strategy satisfies
\begin{align}
 \sup_{t\ge0}\E\Gap(L_t)
 &\le64\alpha_G\frac d\kappa(\log n+D\log4), \label{eq:regular-mean}\\
 \Pr\!\left\{\Gap(L_t)>
       64\alpha_G\frac d\kappa(\log n+D\log4+z)\right\}
 &\le e^{-z}\quad(z\ge0). \label{eq:regular-tail}
\end{align}
The probability estimate holds for every deterministic $t\ge0$.
In particular the expected gap, and the gap with polynomially high
probability at every fixed time, are
$O(\alpha_G(d/\kappa)\log n)$.
Using the decomposition of~\cite{HHR03} gives
$O((d/\kappa)\log^3 n\log\log n)$ as $n\to\infty$.
\end{theorem}

\begin{proof}
The existing flows already provide the inward drift; we verify its
normalization in the original arrival clock.
The edge budget in~\eqref{eq:bf-flows} makes the rule valid.
Since $2M=nd$, summing over the offered edges incident to $v$ gives
\begin{align}
 q_t(v)
 &=\frac1M\sum_{u:\,uv\in E}\frac{1+b_t(u,v)}2
 \nonumber\\
 &=\frac1n-\frac{\beta\kappa}{nd}
                   \sum_I\sgn(\delta_I(t))h_I(v). \label{eq:bf-marginal}
\end{align}
This also recovers the marginal identity in~\cite[Lemma~3.5]{BF22}.
Every valid rule on a regular graph satisfies
$q_t(v)\le d/M=2/n$.
Orthogonality in Lemma~\ref{lem:haar} now yields
\[
 \E[\Delta\delta_I\mid\mathcal F_t]
     =-\frac{\eta\nu_I}{n}\sgn(\delta_I),
 \qquad \eta=\frac{\kappa}{8\alpha_Gd}.
\]
The inequalities $\kappa\le d$ and $\alpha_G\ge1$ imply
$0<\eta\le1/8$. Lemma~\ref{lem:shrinking} gives $B_*\le8$ and
$D=O(\log n)$. Thus every hypothesis of Theorem~\ref{thm:joint} holds.
Substituting $8/\eta=64\alpha_Gd/\kappa$ proves both explicit bounds.
Finally use the congestion guarantee supplied by~\cite{HHR03}
for the construction in~\cite[Section~2.2.1]{BF22}.
\end{proof}

\subsection{Projection through an Euler tour}

Several virtual bins can represent one actual vertex. If their total
capacity matches the actual capacity, summing their loads turns the
normalized actual load into a weighted average. This observation allows
the cycle rule to be used on irregular graphs.

\begin{proposition}\label{prop:projection}
Consider two instances with positive capacities and independent,
identically distributed requests. Let $\pi:\widetilde V\to V$ be surjective, and call $\pi^{-1}(v)$ the
\emph{fiber} over $v$.
Suppose the image of every virtual request is a valid base request,
the push-forward of its law
$\widetilde p$ is $p$, and for some $\lambda>0$,
\[
 \sum_{x:\,\pi(x)=v}\widetilde c_x=\lambda c_v
       \quad(v\in V).
\]
Then every virtual online strategy transfers to a base online strategy
with the pathwise comparison
\[
 H_c(L_t)\le\lambda H_{\widetilde c}(\widetilde L_t)
 \quad\text{at every }t.
\]
Consequently $\OPT(G,p,c)\le\lambda
\OPT(\widetilde G,\widetilde p,\widetilde c)$.
\end{proposition}

\begin{proof}
We lift each request conditionally, so that the virtual process retains
its required independent law. Given a base request $e$ of positive
probability, choose a virtual preimage $\widetilde e$ with conditional
probability $\widetilde p(\widetilde e)/p(e)$, using fresh randomness.
The resulting virtual requests are independent with law $\widetilde p$.
Run the virtual strategy and project its chosen endpoint.
This is online and valid. At every time,
\[
 \frac{L_v}{c_v}
  =\lambda
   \frac{\sum_{\pi(x)=v}\widetilde c_x
                   (\widetilde L_x/\widetilde c_x)}
        {\sum_{\pi(x)=v}\widetilde c_x}.
\]
The fraction is a weighted average of virtual normalized loads.
Taking the range proves the pathwise comparison; taking expectations,
time suprema, and then infima proves the claim about $\OPT$.
\end{proof}

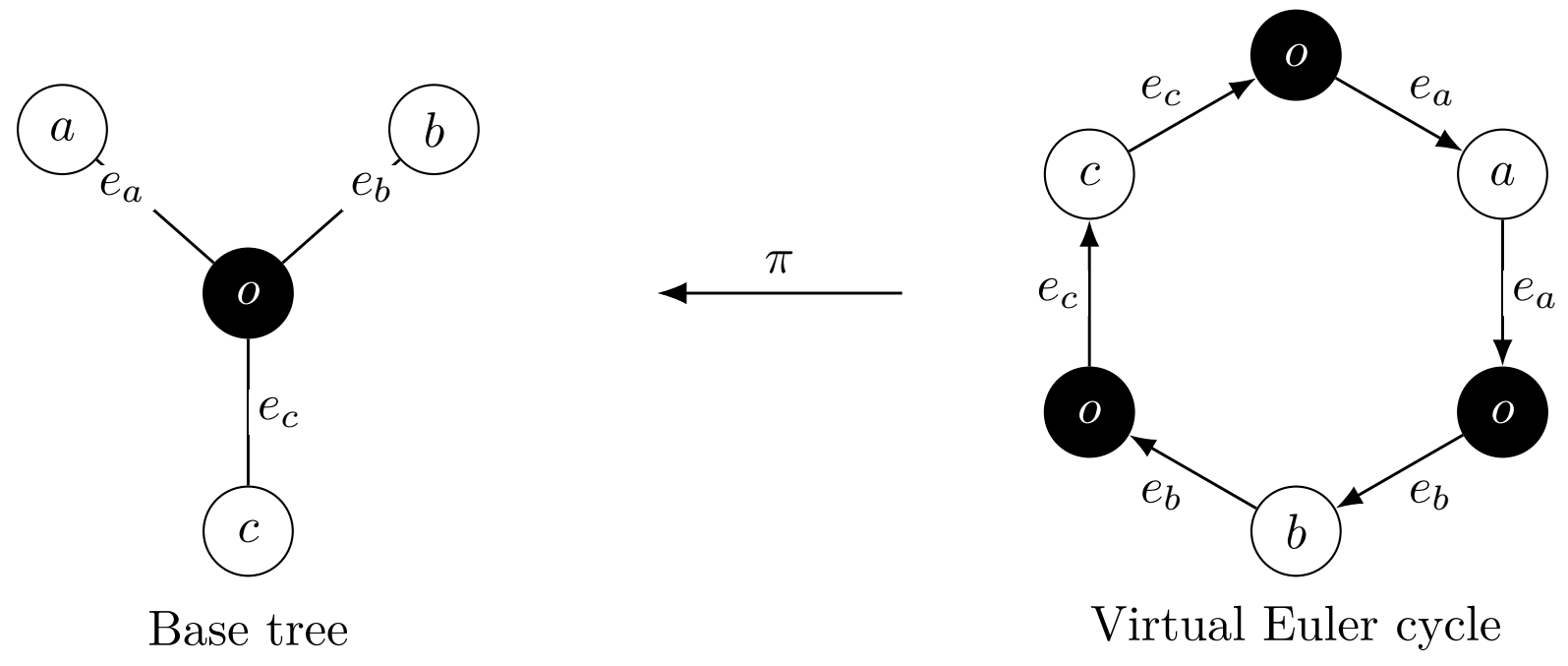
\begin{figure}[htbp]
\centering
\begin{tikzpicture}[
  x=1cm,y=1cm,
  every node/.style={font=\fontsize{9.5}{11}\selectfont},
  vertex/.style={circle,draw,fill=white,minimum size=6mm,inner sep=0pt},
  hub/.style={vertex,fill=black,text=white},
  base edge/.style={line width=.55pt},
  tour edge/.style={line width=.55pt,-{Latex[length=1.8mm,width=1.35mm]}},
  edge label/.style={fill=white,inner sep=1.5pt}
]
  \node[hub] (o) at (-3.8,1.05) {$o$};
  \node[vertex] (a) at (-5.05,2.15) {$a$};
  \node[vertex] (b) at (-2.55,2.15) {$b$};
  \node[vertex] (c) at (-3.8,-.55) {$c$};
  \draw[base edge] (o)--node[edge label,above left] {$e_a$}(a);
  \draw[base edge] (o)--node[edge label,above right] {$e_b$}(b);
  \draw[base edge] (o)--node[edge label,right] {$e_c$}(c);
  \node at (-3.8,-1.2) {Base tree};
  \node[hub] (x0) at (3.25,2.65) {$o$};
  \node[vertex] (x1) at (4.64,1.85) {$a$};
  \node[hub] (x2) at (4.64,.25) {$o$};
  \node[vertex] (x3) at (3.25,-.55) {$b$};
  \node[hub] (x4) at (1.86,.25) {$o$};
  \node[vertex] (x5) at (1.86,1.85) {$c$};
  \draw[tour edge] (x0)--node[edge label,above right] {$e_a$}(x1);
  \draw[tour edge] (x1)--node[edge label,right] {$e_a$}(x2);
  \draw[tour edge] (x2)--node[edge label,below right] {$e_b$}(x3);
  \draw[tour edge] (x3)--node[edge label,below left] {$e_b$}(x4);
  \draw[tour edge] (x4)--node[edge label,left] {$e_c$}(x5);
  \draw[tour edge] (x5)--node[edge label,above left] {$e_c$}(x0);
  \node at (3.25,-1.2) {Virtual Euler cycle};
  \draw[-{Latex[length=2mm]},line width=.55pt]
    (.6,1.05)--node[above] {$\pi$}(-1.05,1.05);
\end{tikzpicture}
\caption{The Euler-tour lift for a three-leaf star. Reading clockwise from
the top gives the virtual vertices \((o,a,o,b,o,c)\).
Every base edge has exactly two virtual-edge preimages, as indicated by
the edge labels. The fiber over \(o\) has size three, and each other fiber
has size one. These fiber sizes are the vertex degrees, which explains
the degree-proportional capacities in the projection.}
\label{fig:euler}
\end{figure}

\begin{proposition}\label{prop:euler}
Let $G$ be a connected simple graph on $n$ vertices with $M\ge2$ edges,
independent uniform requests, and degree-proportional capacities. Then
\begin{equation}\label{eq:euler}
 \OPT(G)\le\dd\,\OPT(C_{2M})
          =O\!\left(\dd\log^2(2M)\right).
\end{equation}
The transferred strategy has the corresponding fixed-time exponential
tail. In particular every tree with degree-proportional capacities has
$\OPT=O(\log^2 n)$.
\end{proposition}

\begin{proof}
We double each edge so that an Euler circuit visits every incidence in
the proportions prescribed by the capacities.
The resulting connected multigraph has even degrees and an Euler circuit.
Its successive $2M$ vertex positions form a virtual cycle $C_{2M}$.
Map each position to the base vertex visited there.
Every base edge has exactly two virtual edge preimages, so choosing a
preimage uniformly lifts uniform base requests to uniform cycle requests.
Vertex $v$ occurs $\deg_G(v)$ times in the circuit positions, because its
indegree in the circuit is half its doubled degree.
Each virtual capacity is one, and therefore its fiber capacity is
$\deg_G(v)=\dd c_v$.
Apply Proposition~\ref{prop:projection} with $\lambda=\dd$, followed by
Theorem~\ref{thm:cycle}. For a tree, $M=n-1$ and $\dd<2$.
The two-vertex tree, excluded by $M\ge2$, has optimal gap one by alternating
its allocations.
\end{proof}

\begin{corollary}\label{cor:path-cycle}
Under uniform edge requests and degree-proportional capacities on paths,
\[
 \OPT(C_n)\le\frac{n+1}{n}\OPT(P_{n+1})\qquad(n\ge3).
\]
Thus a uniform $O(\log n)$ upper bound on paths, on cycles, and on all
trees are equivalent assertions.
\end{corollary}

\begin{proof}
We identify the endpoints of a virtual path to recover a cycle.
Map the vertices $0,\ldots,n$ of $P_{n+1}$ to their residues modulo $n$.
Each path edge maps to a different cycle edge. The two endpoint capacities
are $(n+1)/(2n)$ and every internal capacity is $(n+1)/n$.
Every fiber therefore has total capacity $(n+1)/n$.
Proposition~\ref{prop:projection} gives the displayed inequality.
It transfers a logarithmic path bound to cycles; Proposition~\ref{prop:euler}
transfers a logarithmic cycle bound to trees. Trees include paths.
\end{proof}

\section{Stochastic cut workloads}\label{sec:workload}

\subsection{A universal lower bound}

A request contained in a set $S$ must be allocated there. This forced
work gives a lower bound without specifying a strategy. To retain the
effect of rare overloads, we maximize over the length of the request
prefix before taking expectation.

For a proper nonempty set $S\subset V$, let $N_t(S)$ count the requests
among $e_1,\ldots,e_t$ that are contained in $S$. Put
\begin{align}
 Z_t(S)&=\frac{N_t(S)}{c(S)}
             -\frac{t-N_t(S)}{c(S^c)},&
 R_t(S)&=\pos{Z_t(S)}, \label{eq:cut-walk}\\
 Q&=\sup_{t\ge0}\max_{\varnothing\ne S\subsetneq V}R_t(S),&
 \W&=\frac1{\cmin}+\E Q. \label{eq:workload}
\end{align}
The values $Q$, $\W$, and $\OPT$ may be infinite. The term $1/\cmin$
records the indivisibility of a single ball. It will not be used as a
lower bound for fractional allocations.

\begin{theorem}\label{thm:workload}
For any instance on at least two vertices with positive capacities and
independent identically distributed requests, every online integral
strategy started empty satisfies
\[
 \sup_{t\ge0}\E H_t\ge\frac{\W}{6}.
\]
More precisely, for any nonempty family $\mathcal S$ of proper nonempty sets and
any integer $T\ge0$, let
\[
 Q_T=\max_{0\le s\le T}\max_{S\in\mathcal S}
        \pos{Z_{T-s,T}(S)},\qquad
 K_T=\max_{0\le t\le T}\E H_t,
\]
where $Z_{a,b}$ is~\eqref{eq:cut-walk} formed from the requests in
$(a,b]$. Then $\E Q_T\le4K_T$. This latter inequality also holds for
online fractional allocations.
\end{theorem}

\begin{proof}
We choose a witness window by scanning backwards from a fixed endpoint.
The event selecting a window then depends only on requests after its
starting load, which is where the online property enters.

Define the load contrast
$D_S(t)=L_t(S)/c(S)-L_t(S^c)/c(S^c)$.
If $A$ units are assigned into $S$ during $(a,b]$, then
$A\ge N_{a,b}(S)$. Both normalized averages lie in the range of the
individual normalized loads, so
\begin{equation}\label{eq:window-conservation}
 Z_{a,b}(S)\le D_S(b)-D_S(a)\le H_a+H_b.
\end{equation}
This is pathwise and does not use the online property or integrality.

Fix $T\ge1$ and write $K=K_T$.
If $K=0$, every $H_t$ with $t\le T$ vanishes almost surely and
\eqref{eq:window-conservation} gives $Q_T=0$.
Otherwise, for $x\ge0$, define
\[
 \tau_x=\min\{1\le s\le T:
              \max_{S\in\mathcal S}Z_{T-s,T}(S)>x\},
\]
with value $\infty$ when this set is empty.
For each fixed $s$, the event $\{\tau_x=s\}$ depends only on
$e_{T-s+1},\ldots,e_T$ and is independent of $H_{T-s}$.
For $u>0$, Markov's inequality therefore gives
\begin{align}
 \Pr(\tau_x\le T,\ H_{T-\tau_x}>u)
 &=\sum_{s=1}^T\Pr(\tau_x=s)\Pr(H_{T-s}>u)
 \nonumber\\
 &\le\frac Ku\Pr(\tau_x\le T). \label{eq:reverse-independence}
\end{align}
The random variable $H_{T-\tau_x}$ is used only on $\{\tau_x\le T\}$.
On that event,~\eqref{eq:window-conservation} implies
$x<H_T+H_{T-\tau_x}$.
Taking $u=2K$ in~\eqref{eq:reverse-independence} yields
\[
 \Pr(Q_T>x)\le2\Pr(H_T>x-2K)\qquad(x>2K).
\]
Tail integration proves
$\E Q_T\le2K+2\E H_T\le4K$.
The case $T=0$ has both sides zero.

Reversal preserves the law of a finite independent identically
distributed request sequence. Thus $Q_T$ has the same distribution as
$\max_{0\le t\le T}\max_{S\in\mathcal S}R_t(S)$.
These forward maxima increase with $T$. Monotone convergence, with all
proper nonempty sets as $\mathcal S$, gives
\begin{equation}\label{eq:cut-necessary}
 \E Q\le4\sup_{t\ge0}\E H_t.
\end{equation}

For the remaining indivisibility term choose $v$ with $c_v=\cmin$,
write $\rho=c_v/C\le1/2$, and let $t$ be an integer nearest to
$1/(2\rho)$. Then
$|t\rho-1/2|\le\rho/2$ and
$\operatorname{dist}(t\rho,\mathbb Z)\ge(1-\rho)/2$.
Since $L_t(v)$ is an integer,
\[
 H_t\ge
 \left|\frac{L_t(v)}{c_v}-\frac{t-L_t(v)}{C-c_v}\right|
 =\frac{|L_t(v)-t\rho|}{c_v(1-\rho)}
 \ge\frac1{2\cmin}.
\]
This witness time depends only on the capacities.
Adding this bound to~\eqref{eq:cut-necessary} proves the theorem,
including the case of infinite cost.
\end{proof}

\begin{remark}\label{rem:workload-projection}
Under the coupling of Proposition~\ref{prop:projection}, a base cut $S$
and its virtual preimage have identical forced counts, while their
capacity masses differ by $\lambda$. Hence
$R_t(S)=\lambda\widetilde R_t(\pi^{-1}S)$.
Also $\lambda c_v\ge\widetilde c_{\min}$ for every $v$.
It follows that $\W(G,p,c)\le\lambda
\W(\widetilde G,\widetilde p,\widetilde c)$.
\end{remark}

\subsection{Comparison with separately optimized batches}

For a fixed batch, a flow calculation characterizes feasible load
intervals. This is the classical orientation and submodular feasibility
setting of Hakimi~\cite{Hakimi65} and Fujishige~\cite{Fujishige05}.
We give the calculation to specify the weighted gap and the role of
reassignment exactly. The static feasibility principle is not a new
ingredient.

\begin{proposition}\label{prop:offline}
Fix a batch of $t\ge0$ nonempty requests of size at most two and
positive vertex capacities. For a nonempty set $S\subseteq V$ let
$N(S)$ count requests contained in $S$ and $M(S)$ requests touching $S$.
Define
\[
 U=\max_{\varnothing\ne S\subseteq V}\frac{N(S)}{c(S)},\qquad
 A=\min_{\varnothing\ne S\subseteq V}\frac{M(S)}{c(S)},\qquad
 R=\max_{\varnothing\ne S\subsetneq V}R_t(S).
\]
If $\OFF_{\mathrm{frac}}$ and $\OFF$ are the minimum fractional and
integral final gaps of this batch, then
\begin{align}
 \OFF_{\mathrm{frac}}&=U-A,&
 U-A\le\OFF&\le U-A+\frac2{\cmin}, \label{eq:offline-formula}\\
 R&\le\OFF_{\mathrm{frac}}\le2R,&
 R\le\OFF&\le2R+\frac2{\cmin}. \label{eq:offline-cut}
\end{align}
\end{proposition}

\begin{proof}
We prove simultaneous lower and upper load constraints, since optimizing
the two endpoints of the gap separately would not suffice.
For vectors $0\le\ell\le u$, a fractional load vector in the box
$[\ell,u]$ exists precisely when
\begin{equation}\label{eq:box-cuts}
 N(S)\le u(S),\qquad \ell(S)\le M(S)
                    \quad(S\subseteq V).
\end{equation}
Necessity follows from forced and available jobs.
For sufficiency form a circulation network with a source $s$, one node
per job, one node per vertex, and a sink $z$.
The arc $s\to\text{job}$ has lower and upper capacity one.
Each job has an infinite-capacity arc to each eligible vertex.
The arc $v\to z$ has bounds $\ell_v,u_v$, and $z\to s$ has both
bounds $t$.

The circulation cut criterion requires that the total lower capacity
entering each node set is at most its total upper capacity leaving.
It follows from max-flow/min-cut by subtracting lower capacities and
adding a super-source and super-sink for the resulting imbalances.
For completeness, a finite cut here consists of some job nodes $J$ and
vertex nodes $S$, with all endpoints of jobs in $J$ contained in $S$;
thus $|J|\le N(S)$.
If the cut contains neither $s$ nor $z$, or contains $s$ but not $z$,
its inequality reduces to $|J|\le u(S)$.
If it contains both, or contains $z$ but not $s$, the inequality reduces
to $\ell(S^c)\le t-|J|$.
Both follow from~\eqref{eq:box-cuts}, since
$M(S^c)=t-N(S)$. Cuts of infinite upper capacity impose no constraint.
This proves sufficiency. With integer bounds the same reduction has an
integral maximum flow, so the box criterion also gives an integral
allocation.

Every allocation has maximum normalized load at least $U$ and minimum
at most $A$. The full set gives $A\le t/C\le U$.
The box $\ell_v=Ac_v$, $u_v=Uc_v$ satisfies~\eqref{eq:box-cuts},
which proves the equality in~\eqref{eq:offline-formula}.
If $y$ is a feasible fractional load vector in this box, it is feasible
also in $[\lfloor y\rfloor,\lceil y\rceil]$.
The integral box criterion gives an integral load vector in that box;
normalizing changes each endpoint by at most $1/\cmin$.

Finally, for a proper set $S$,
\[
 \frac{N(S)}{c(S)}-\frac tC
 =\frac{c(S^c)}C
     \left(\frac{N(S)}{c(S)}
              -\frac{t-N(S)}{c(S^c)}\right).
\]
Hence $U-t/C\le R$.
The identity $M(S)=t-N(S^c)$ gives $t/C-A\le R$.
Conversely, $U\ge N(S)/c(S)$ and
$A\le M(S^c)/c(S^c)$ imply $U-A\ge R$.
Combine these inequalities with~\eqref{eq:offline-formula}.
\end{proof}

The proposition allows every prefix to be allocated afresh. It therefore
does not produce a single irrevocable trajectory.
This distinction will matter in Theorem~\ref{thm:hub}.

\begin{corollary}\label{cor:batch}
For the independent, identically distributed request model with positive
capacities, let
$\mathcal B=\E\sup_{t\ge0}\OFF(e_1,\ldots,e_t;c)$, where each prefix is
optimized separately. Then
\[
 \frac{\W}{3}\le\mathcal B\le2\W,\qquad \OPT\ge\frac{\mathcal B}{4}.
\]
\end{corollary}

\begin{proof}
The batch inequalities compare the benchmarks, while the reverse-window
argument compares a batch with any online trajectory.
Equation~\eqref{eq:offline-cut} gives
$\mathcal B\ge\E Q$ and $\mathcal B\le2\W$.
The deterministic witness time for indivisibility in the proof of
Theorem~\ref{thm:workload} applies to every integral batch allocation,
so $\mathcal B\ge1/(2\cmin)$ and $\W\le3\mathcal B$.

For an online trajectory, its assignments during $(a,b]$ form a feasible
batch allocation. The range of its normalized increment vector is at
most $H_a+H_b$, so
$\OFF(e_{a+1},\ldots,e_b;c)\le H_a+H_b$.
The event selecting the first backwards suffix with offline optimum
greater than $x$ depends only on that suffix's requests.
It is therefore independent of its starting load.
The argument in~\eqref{eq:reverse-independence} bounds the expected
largest suffix optimum by $4\max_{t\le T}\E H_t$.
Time reversal and monotone convergence give
$\mathcal B\le4\sup_t\E H_t$, and taking the infimum proves the claim.
\end{proof}

\section{Cut tails and many bottlenecks}\label{sec:bottlenecks}

\subsection{A uniform tail for degree-proportional capacities}

We first record the exponential crossing estimate used in the cut
arguments. It controls an infinite sequence of prefixes directly.

\begin{lemma}\label{lem:exponential-max}
Let $\xi_1,\xi_2,\ldots$ be independent with a common law. If
$\theta>0$ and $\E e^{\theta\xi_1}\le1$, then
\[
 \Pr\!\left\{\sup_{t\ge0}\sum_{j=1}^t\xi_j>x\right\}
      \le e^{-\theta x}\qquad(x\ge0).
\]
\end{lemma}

\begin{proof}
We stop the exponential before its first crossing.
The process $\exp(\theta\sum_{j=1}^t\xi_j)$ is a nonnegative
supermartingale starting at one.
Stopping at the smaller of a finite horizon and the first crossing
of $x$ bounds the crossing probability by $e^{-\theta x}$.
Letting the horizon increase proves the result by continuity from below.
\end{proof}

\begin{proposition}\label{prop:cut-upper}
For a connected simple graph on $n\ge2$ vertices with independent
uniform edge requests and degree-proportional capacities, write $\kappa$ for its edge
connectivity and $b_S=|\partial S|$. Then
\begin{align}
 \Pr\{\sup_t R_t(S)>x\}
       &\le\exp\!\left(-\frac{b_S}{2\dd}x\right), \label{eq:one-cut-tail}\\
 \W(G)&\le\frac{\dd}{\kappa}\bigl(8\log(2n)+5\bigr).
       \label{eq:all-cut-upper}
\end{align}
\end{proposition}

\begin{proof}
Natural capacities express the drift of a cut walk directly in terms
of its boundary; cut counting then pays for the maximum over sets.
Put $a=c(S)$, $b=c(S^c)$ and $\gamma=b_S/\dd$.
The degree-sum identity gives
\[
 a+b=n,\qquad 0<\gamma\le\min(a,b),\qquad
 p_S:=\Pr(e_1\subseteq S)=\frac{a-\gamma}{n}.
\]
A cut increment $\xi$ equals $1/a$ with probability $p_S$ and
$-1/b$ otherwise. Its first two moments satisfy
\[
 \E\xi=-\frac{\gamma}{ab},\qquad
 ab\,\E\xi^2=1+\frac{\gamma}{n}
                     \left(\frac ab-\frac ba\right)\le2.
\]
If $a\le b$ the correction is nonpositive; otherwise use $\gamma\le b$
to bound it by $a/n\le1$.
For $\theta=\gamma/2$, we have $|\theta\xi|\le1/2$, so
\[
 \E e^{\theta\xi}
 \le1+\theta\E\xi+\theta^2\E\xi^2
 \le1+\frac{-\theta\gamma+2\theta^2}{ab}=1.
\]
Lemma~\ref{lem:exponential-max} proves~\eqref{eq:one-cut-tail}.

We use the following coarse version of the random-contraction
cut-counting argument of Karger~\cite{Karger00}: the number of oriented
cuts of size at most $j\kappa$ is at most $(2n)^{2j}$ for every integer
$j\ge1$. Here an oriented cut means a choice of $S$, so complements
are counted separately.
If $2j\le n$, contract random edges until $2j$ vertices remain.
Conditional on a fixed such cut surviving at $v$ vertices, the contracted
multigraph has minimum cut at least $\kappa$ and hence at least
$v\kappa/2$ edges. The chance of avoiding the cut on the next step is
at least $1-2j/v$.
Its survival probability is at least
\[
 \prod_{v=2j+1}^n\left(1-\frac{2j}{v}\right)
       =\binom n{2j}^{-1}.
\]
Choosing a uniform subset of the surviving vertices selects this
oriented cut with conditional probability $2^{-2j}$.
Since output probabilities sum to at most one, the number of cuts is
at most $2^{2j}\binom n{2j}\le(2n)^{2j}$.
For $2j>n$, the bound follows from the total count $2^n$.

Put $\lambda=\kappa/(2\dd)$ and $a_0=2n$.
Grouping cuts by $j\kappa\le b_S<(j+1)\kappa$ gives
\[
 \Pr(Q>x)\le\sum_{j\ge1}a_0^{2(j+1)}e^{-j\lambda x}.
\]
At $x_0=\lambda^{-1}(4\log a_0+\log2)$, the geometric series implies
$\Pr(Q>x_0+y)\le e^{-\lambda y}$ for $y\ge0$:
its ratio is at most $1/(2a_0^2)$ and its first term is
$e^{-\lambda y}/2$.
Thus $\E Q\le x_0+\lambda^{-1}$.
Finally $1/\cmin=\dd/\delta(G)\le\dd/\kappa$.
Adding this term and using $2\log2+3<5$ proves~\eqref{eq:all-cut-upper}.
\end{proof}

\subsection{Disjoint bottlenecks}

One weak cut produces an exponential excursion scale. Many disjoint weak
cuts increase their maximum by a logarithmic factor. Their request
streams compete for arrivals, but that competition has a favorable
direction for the second-moment estimate.

\begin{lemma}\label{lem:disjoint}
In the independent, identically distributed request model with positive
capacities, let $S_1,\ldots,S_N$ be pairwise disjoint proper nonempty sets,
with $0<p_i=\Pr(e_1\subseteq S_i)<c(S_i)/C$.
Let $\theta_i>0$ be the positive solution of
\[
 p_i e^{\theta_i/c(S_i)}
       +(1-p_i)e^{-\theta_i/c(S_i^c)}=1.
\]
For $x>0$, put
\[
 m(x)=\sum_{i=1}^N
       \exp\!\left[-\theta_i\left(x+\frac1{c(S_i)}\right)\right].
\]
Then
\begin{equation}\label{eq:many-cut-tail}
 \Pr\{\max_i\sup_t R_t(S_i)>x\}\ge\frac{m(x)}{1+m(x)},
 \qquad
 \OPT\ge\frac x4\,\frac{m(x)}{1+m(x)}.
\end{equation}
\end{lemma}

\begin{proof}
We combine a one-cut hitting estimate with pairwise negative correlation,
which is enough to control the probability that at least one cut hits.
For a fixed $i$, the increment moment generating function is strictly
convex, has negative derivative at zero, and diverges at positive
infinity. Its positive root therefore exists uniquely.
Let $\tau$ be the first time the untruncated cut walk exceeds $x$.
The stopped martingale $e^{\theta_i Z_{t\wedge\tau}(S_i)}$ is bounded
by $e^{\theta_i(x+1/c(S_i))}$.
On $\{\tau=\infty\}$ the walk tends to negative infinity by the strong
law. Dominated convergence gives
\[
 1=\E[e^{\theta_i Z_\tau(S_i)};\tau<\infty]
 \le e^{\theta_i(x+1/c(S_i))}\Pr(\tau<\infty).
\]

To prove negative correlation, first restrict to a finite horizon.
For $i\ne j$, condition on all indicators
$I_s=\1_{\{e_s\subseteq S_i\}}$ in that horizon.
Conditional on this vector, the indicators for $S_j$ are independent
over time, with probability zero when $I_s=1$ and
$p_j/(1-p_i)$ when $I_s=0$.
The event that the $S_i$ walk hits $x$ is increasing in $(I_s)$.
The conditional probability that the $S_j$ walk hits $x$ is decreasing
in the same vector.

By the product-measure correlation inequality of Harris~\cite{Harris60},
an increasing and a decreasing function of independent Bernoulli
coordinates have nonpositive covariance. Here is a direct verification:
for one coordinate the covariance is
$p(1-p)(f(1)-f(0))(g(1)-g(0))\le0$.
Conditioning on one coordinate and using the covariance decomposition
proves the assertion inductively, because conditional means preserve
the two monotonicity directions.
The finite-horizon hitting events are therefore negatively correlated.
Increasing the horizon preserves the inequality.

Let $Y$ count the cuts whose workloads ever exceed $x$. The hitting
estimate gives $\E Y\ge m(x)$, and negative correlation gives
$\E Y^2\le(\E Y)^2+\E Y$.
Cauchy--Schwarz yields
\[
 \Pr(Y>0)\ge\frac{(\E Y)^2}{\E Y^2}
          \ge\frac{m(x)}{1+m(x)}.
\]
The second assertion of~\eqref{eq:many-cut-tail} follows from
$\E Q\ge x\Pr(Y>0)$ and~\eqref{eq:cut-necessary}.
\end{proof}

\begin{figure}[tb]
\centering
\begin{tikzpicture}[font=\small,scale=.93]
 \foreach \j/\x/\y in {1/0/1.3,2/4.4/1.3,3/4.4/-1.4,4/0/-1.4}{
   \node[draw,rounded corners,minimum width=2.4cm,minimum height=1.3cm]
       (Q\j) at (\x,\y) {};
   \node at (\x,\y+.24) {$Q_{\j}$};
   \node at (\x,\y-.22) {$K_r\setminus\text{$k$ edges}$};
 }
 \draw[thick] (Q1.east)--node[above]{$k$}(Q2.west);
 \draw[thick] (Q2.south)--node[right]{$k$}(Q3.north);
 \draw[thick] (Q3.west)--node[below]{$k$}(Q4.east);
 \draw[thick] (Q4.north)--node[left]{$k$}(Q1.south);
 \node[align=left,anchor=west] at (6.1,.25)
 {$d=r-1$\\[3pt]$|\partial Q_j|=2k$\\[3pt]$\kappa=2k$};
\end{tikzpicture}
\caption{The ring construction, shown with four blocks. In each block the
deleted edges form a matching. Their endpoints are joined to the next
block by $k$ new edges, restoring every degree to $r-1$.
Each line between blocks represents these $k$ edges.}
\label{fig:ring}
\end{figure}
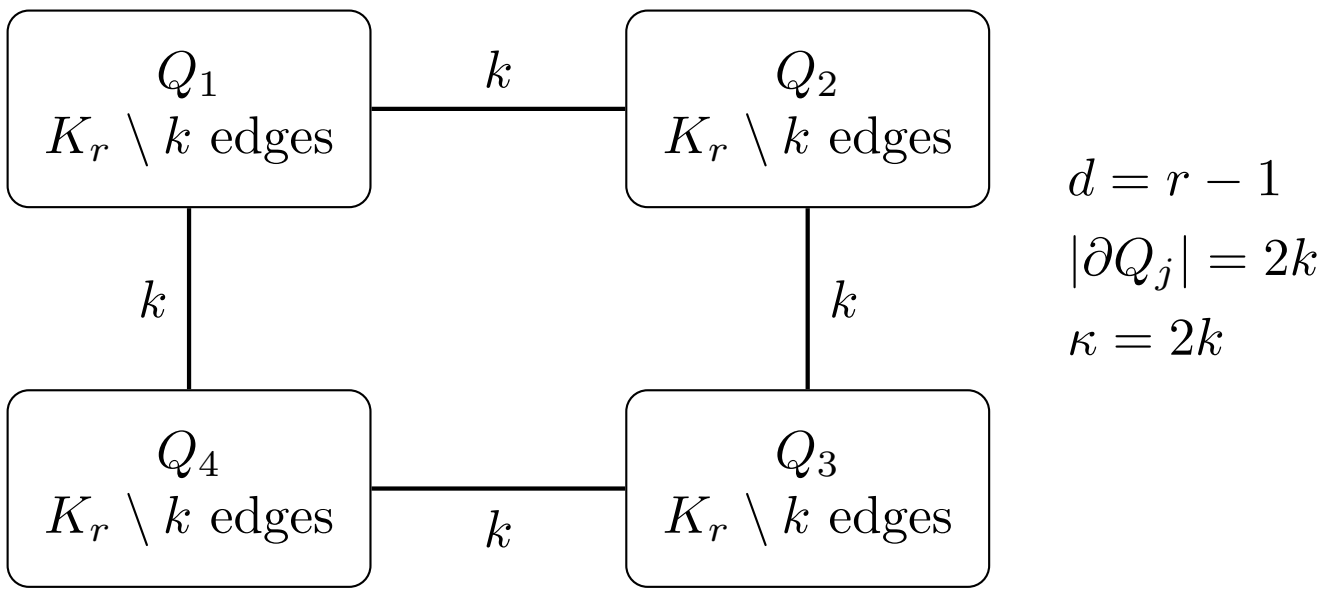

\begin{theorem}\label{thm:rings}
Let $N\ge3$, $r\ge4$, and $1\le k\le\lfloor(r-2)/2\rfloor$.
From each of $N$ disjoint copies $Q_j$ of $K_r$, delete a matching
$\{a_{j,i}b_{j,i}:1\le i\le k\}$. Add the edges
$b_{j,i}a_{j+1,i}$, with block indices modulo $N$.
The resulting simple graph has
\[
 n=Nr,\qquad d=r-1,\qquad\kappa=2k.
\]
With independent uniform edge requests, unit capacities, and $L_0=0$,
\begin{align}
 \OPT(G)&\ge\frac{d}{256k}\log N
              =\frac{d}{128\kappa}\log N, \label{eq:ring-opt}\\
 \frac{d}{64k}\log N
       &\le\W(G)\le\frac{d}{2k}\bigl(8\log(2n)+5\bigr).
       \label{eq:ring-workload}
\end{align}
In particular $N=r$ and $k=1$ give $\OPT(G)=\Omega(\sqrt n\log n)$.
\end{theorem}

\begin{proof}
We retain exact regularity by replacing each deleted incidence with one
interblock incidence. Every vertex has degree $r-1$, and every block has
boundary size $2k$. If a cut splits a block, and the smaller part of that
split has $a$ vertices, at least
\[
 a(r-a)-a=a(r-a-1)\ge r-2\ge2k
\]
internal edges cross. The middle inequality holds for
$1\le a\le r/2$: the quadratic is concave, and its values at the two
endpoints are at least $r-2$.
If no block is split, a nontrivial union of blocks crosses at least two
interfaces, each containing $k$ edges. Thus $\kappa=2k$.

For one block put $\varepsilon=2k/(r(r-1))\le1/4$.
Its forced probability is $p=(1-\varepsilon)/N$ and its workload
increments are $1/r$ and $-1/[r(N-1)]$.
If $\theta$ is its positive exponential root and $u=\theta/r$, its
root equation is
\[
 f(u)=p e^u+(1-p)e^{-u/(N-1)}=1.
\]
For $u\ge0$, the inequalities $e^u\ge1+u+u^2/2$ and
$e^{-v}\ge1-v$ give
\[
 f(u)\ge1-\frac{\varepsilon}{N-1}u+\frac p2u^2.
\]
Since $2(1-\varepsilon)(N-1)\ge N$, we have $f(4\varepsilon)\ge1$.
Strict convexity and the negative derivative at zero imply
\[
 \theta\le4r\varepsilon=\frac{8k}{d},
 \qquad \frac{\theta}{r}\le1.
\]
Set $x=(d/(16k))\log N$. Lemma~\ref{lem:disjoint} gives
$m(x)\ge e^{-1}\sqrt N\ge e^{-1}\sqrt3>1/3$.
Consequently $m(x)/(1+m(x))\ge1/4$.
Thus $\E Q\ge x/4$ and $\OPT\ge x/16$, which give the lower
bounds. Proposition~\ref{prop:cut-upper} gives the upper bound.
For $N=r$, the identities $n=r^2$ and $d=r-1$ give the final assertion.
\end{proof}

\subsection{Trees}

For trees the cut workload itself has logarithmic order. The lower
bound uses many low-degree vertices that can remain unoffered; the
upper bound is the degree-proportional cut estimate.

\begin{corollary}\label{cor:tree-workload}
For every tree $T$ on $n\ge2$ vertices, with independent uniform edge
requests and degree-proportional capacities,
\[
 \frac1{100}\log n\le\W(T)\le16\log(2n)+10.
\]
Consequently
$\OPT(T)\ge(\log n)/600$, while
Proposition~\ref{prop:euler} gives $\OPT(T)=O(\log^2 n)$.
\end{corollary}

\begin{proof}
We select nonadjacent low-degree vertices so that their missing-offer
events have nonpositive pairwise covariance.
The upper bound follows from $\dd<2$ and $\kappa=1$.
For $n\ge8$, the degree sum $2n-2$ implies that at least $n/2$
vertices have degree at most three. Their induced forest has an
independent set $I$ of size at least $n/4$.
Put $t=\lfloor(n-1)\log n/12\rfloor$ and let $Y$ count vertices of
$I$ not offered in the first $t$ requests.
A vertex $v\in I$ has offer probability
$p_v\le3/(n-1)<1/2$. Using
$\log(1-z)\ge-z/(1-z)$ gives
\[
 \Pr(v\text{ unoffered})\ge e^{-6t/(n-1)}\ge n^{-1/2},
 \qquad \E Y\ge\frac{\sqrt n}{4}.
\]
The incident edge sets of two vertices in $I$ are disjoint, so their
joint missing probability is
$(1-p_u-p_v)^t\le(1-p_u)^t(1-p_v)^t$.
Hence $\E Y^2\le(\E Y)^2+\E Y$ and
$\Pr(Y>0)\ge\E Y/(1+\E Y)\ge1/3$.
If $v$ is unoffered, all requests lie in $V\setminus\{v\}$, whose
workload is at least $t/n$.
The unfloored value defining $t$ is at least one, so
$t/n\ge7\log n/192$.
Thus $\W(T)\ge7\log n/576\ge(\log n)/100$.
For $2\le n<8$, the term $1/\cmin=\dd\ge1$ suffices.
Apply Theorem~\ref{thm:workload} for the gap lower bound.
\end{proof}

\section{Heterogeneous capacities: cut workloads do not determine the gap}
\label{sec:heterogeneous}

The workload in \eqref{eq:workload} measures assignments that a request
sequence forces into a set.  With heterogeneous capacities, this information
does not determine the cost of maintaining a single allocation trajectory.
We give two examples on complete support graphs.  In the first,
$\W$ remains bounded while every irrevocable trajectory has expected gap
of order at least $\log m$, even if assignments may be fractional and the
entire request sequence is known in advance (that is, even for
\emph{clairvoyant} trajectories).  In the second, the relative
slack of every cut is bounded below by a constant, yet every online
integral strategy incurs a gap of order at least $\log\log m$.

The budget arguments below allow irrevocable trajectories chosen using
the entire infinite request sequence.  The requests themselves retain
their specified independent, identically distributed law.  All trajectories
start from zero.

\subsection{Missing offers and the budget for external assignments}

Fix a nonempty proper set $S\subset V$, and write
\[
 C=c(V),\qquad
 \Delta_S=\frac{c(S)}C-\Pr(e\subseteq S)>0.
\]
Every request contained in $S$ must be assigned into $S$.  Consequently,
if the mean normalized load of $S$ is to follow the global mean,
its long-run allowance for assignments from other requests is only
$\Delta_S$ per step.  This allowance can be small even when such external
requests are offered frequently.

For integers $u\ge0$ and $T\ge1$, let $Y_{u,T}$ count the vertices of $S$
that receive no offer from a request contained in $S$ during $(u,u+T]$.
Write $Y_T=Y_{0,T}$.  The i.i.d. request law gives
$\Pr(Y_{u,T}\ge z)=\Pr(Y_T\ge z)$ for every $u$ and $z$.

\begin{lemma}[Integral correction budget]\label{lem:budget}
For the i.i.d. request model with positive capacities, fix a nonempty
proper $S$ with $\Delta_S>0$ and an integer $T\ge1$, as above.
For every irrevocable integral trajectory starting empty and every $z>0$,
\begin{equation}\label{eq:integral-budget}
 \sup_{t\ge0}\E H_t
 \ge \frac{T}{2C}
 \left[\Pr(Y_T\ge z)-\frac{T\Delta_S}{z}\right]_+.
\end{equation}
The trajectory need not be online.
\end{lemma}

\begin{proof}
A vertex missed by the internal stream must receive an external
assignment if its normalized load is not to fall behind during the
window.  Put $h=T/C$, and let $B_u$ be the number of assignments into $S$
from requests not contained in $S$ during $(u,u+T]$.  On the event
$Y_{u,T}\ge z$ and $B_u<z$, at least one of the internally unoffered vertices
receives no job at all.

By \eqref{eq:centered-range}, $|X_v(t)|\le H_t$ for every $v$.  A vertex receiving no job during the
window satisfies $X_v(u+T)-X_v(u)=-h$.  Hence
$H_u+H_{u+T}\ge h$ on the event just described.  Markov's inequality
gives
\begin{equation}\label{eq:integral-budget-window}
 \E H_u+\E H_{u+T}
 \ge h\left(\Pr(Y_T\ge z)-\frac{\E B_u}{z}\right).
\end{equation}
This argument requires no independence between $B_u$, the missed
vertices, and the loads.

Let $K=\sup_t\E H_t$.  There is nothing to prove if $K=\infty$.
Otherwise, write the centered total load of $S$ as
\[
 X_S(t)=\sum_{v\in S}c_vX_v(t)=L_t(S)-tc(S)/C.
\]
Conservation over $r$
consecutive windows, together with the empty initial state, gives
\begin{equation}\label{eq:external-budget-balance}
 \E\sum_{j=0}^{r-1}B_{jT}
 =rT\Delta_S+\E X_S(rT).
\end{equation}
Indeed, the total assignments into $S$ equal the number of internal
requests plus the number counted by the left side.  Also
$|\E X_S(rT)|\le c(S)K$.  Sum
\eqref{eq:integral-budget-window}, divide by $r$, and let $r\to\infty$.
The left side is at most $2K$, and the resulting inequality is
\eqref{eq:integral-budget}.
\end{proof}

For fractional allocations, counting external assignments is
insufficient: a request may contribute an arbitrarily small amount.
Counting the mass needed to repair the missed vertices gives the
following extension.

\begin{lemma}[Fractional correction budget]\label{lem:fractional-budget}
For the i.i.d. request model with positive capacities, fix a nonempty
proper $S$ with $\Delta_S>0$ and an integer $T\ge1$, as above,
and put $c_*(S)=\min_{v\in S}c_v$. Every irrevocable fractional
trajectory starting empty satisfies, for every $z>0$,
\begin{equation}\label{eq:fractional-budget}
 \sup_{t\ge0}\E H_t
 \ge\frac{T}{4C}
 \left[\Pr(Y_T\ge z)-\frac{2C\Delta_S}{c_*(S)z}\right]_+.
\end{equation}
The trajectory may depend on the entire infinite request sequence.
\end{lemma}

\begin{proof}
Keeping the gap small at both ends of a window forces a definite amount
of external mass into every internally unoffered vertex.  Again put
$h=T/C$, and now let $B_u$ be the total mass assigned into $S$ from
requests not contained in $S$ during $(u,u+T]$.  If
$H_u+H_{u+T}<h/2$, then for every vertex $v$,
\[
 \frac{L_{u+T}(v)-L_u(v)}{c_v}
 =h+X_v(u+T)-X_v(u)>h/2.
\]
If in addition $Y_{u,T}\ge z$, at least $z$ vertices must receive all this
mass externally.  Thus $B_u\ge c_*(S)hz/2$.  It follows that
\[
 \E H_u+\E H_{u+T}
 \ge\frac h2\Pr(Y_T\ge z)-\frac{\E B_u}{c_*(S)z}.
\]
For finite $K=\sup_t\E H_t$, sum over $r$ consecutive windows and use
\eqref{eq:external-budget-balance}, which remains valid for mass.
After division by $r$ and passage to the limit,
\[
 2K\ge\frac h2\Pr(Y_T\ge z)-\frac{T\Delta_S}{c_*(S)z}.
\]
This is \eqref{eq:fractional-budget}.  The case $K=\infty$ is immediate.
Neither the conservation identity nor the event inclusion uses
the online property.
\end{proof}

\subsection{A logarithmic obstruction that survives clairvoyance}

For an integer $m\ge16$, consider $K_{m+1}$ with ordinary vertices
$[m]$ of capacity one and a hub $0$ of capacity $m^2$.  Write
$C=m^2+m$.  Give each ordinary pair probability $2/(mC)$ and each hub
pair probability
\[
 p_*:=\frac{1-(m-1)/C}{m}.
\]
The total probability of an ordinary request is $(m-1)/C$, whereas the
capacity share of the ordinary set is $m/C$.  Thus, for $S=[m]$,
\begin{equation}\label{eq:hub-slack}
 \Delta_S=1/C.
\end{equation}
This law is strictly feasible: for an ordinary set $A$ of size $s$,
\[
 q_A=\Pr(e\subseteq A)=\frac{s(s-1)}{mC}<\frac{s}{C}.
\]
For a cut containing the hub, write its complement as an ordinary set
$B$ of size $r$.  A request touches $B$ with probability at least
$rp_*$.  Since $Cp_*=m+1/m>1$, this probability exceeds $r/C$.
Equivalently, the probability of a request wholly in $V\setminus B$
is less than $c(V\setminus B)/C$.  These cases exhaust the proper cuts.
All edge probabilities are positive and sum to one.

\begin{lemma}[Missing internal offers]\label{lem:hub-coupons}
For the independently sampled near-critical family above with $m\ge16$,
let
\[
 h_0=\frac{\log m}{16},\qquad T=\lfloor Ch_0\rfloor,
 \qquad h=T/C,
\]
and let $Y_T$ count the ordinary vertices receiving no internal offer in
a length-$T$ window.  Then
\[
 h\ge\frac{\log m}{32},\qquad
 \mu:=\E Y_T\ge m^{3/4},\qquad
 \Pr(Y_T\ge\mu/2)\ge\frac18.
\]
\end{lemma}

\begin{proof}
The internal stream offers any one vertex at rate at most $2/C$, so a
window of length $C\log m/16$ still misses a polynomial number of
vertices; a two-vertex calculation controls their dependence.
Since $C\log m/32\ge1$ for $m\ge16$, flooring gives
$h\ge\log m/32$.  The probability that a given ordinary
vertex is offered in an internal request is
\[
 \rho=\frac{2(m-1)}{mC}\le\frac2C.
\]
Consequently, with $Y_T$ as in the budget lemmas,
\[
 \mu:=\E Y_T=m(1-\rho)^T\ge m^{3/4}.
\]
Indeed, $-\log(1-\rho)\le\rho/(1-\rho)$ and
\[
 \frac{T\rho}{1-\rho}
 \le\frac{2Ch_0}{C-2}\le4h_0=\frac14\log m.
\]

Two distinct ordinary vertices are jointly offered by their common
pair, whose probability is $p=2/(mC)$.  Thus their joint probability
of being internally unoffered is $(1-2\rho+p)^T$.  Since
$(1-\rho)^2\ge1/2$,
\[
 \left(\frac{1-2\rho+p}{(1-\rho)^2}\right)^T
 \le\exp\left(\frac{Tp}{(1-\rho)^2}\right)
 \le e^{4h_0/m}.
\]
It follows that
\begin{equation}\label{eq:hub-coupon-moments}
 \frac{\E Y_T^2}{\mu^2}
 \le\frac1\mu+e^{4h_0/m}\le2,
 \qquad
 \Pr(Y_T\ge\mu/2)\ge\frac18.
\end{equation}
For the first inequality's numerical bound, use
$\mu\ge m^{3/4}\ge8$ and
$4h_0/m=\log m/(4m)\le\log16/64<\log(3/2)$.
The probability bound follows from the second-moment inequality
$\Pr(Y\ge\E Y/2)\ge(\E Y)^2/(4\E Y^2)$.
\end{proof}

Write $N_t(A)$ for the number of requests contained in an ordinary set
$A$, put $N_t=N_t([m])$, and let $M_t(B)$ count internal ordinary
requests touching $B$.  Define
\[
\begin{aligned}
 Z_{\mathrm{small}}
   &=\sup_t\max_{2\le|A|\le m/2}
      [N_t(A)/|A|-t/C]_+,\\
 F&=\sup_t[N_t/m-t/C]_+,\\
 Z_{\mathrm{miss}}&=\sup_t\max_{1\le|B|\le m/2}
      [|B|t/C-M_t(B)]_+,\\
 Z&=\sup_t\max_{\varnothing\ne A\subseteq[m]}
      [N_t(A)/|A|-t/C]_+.
\end{aligned}
\]

These variables are defined on the same infinite request sequence.
The following estimates isolate the three fluctuation bounds needed for
the joint cut calculation.

\begin{lemma}[Internal workload estimates]\label{lem:hub-internal}
For the independently sampled near-critical family with $m\ge16$,
\[
 \E Z_{\mathrm{small}}<5,\qquad \E F\le1,\qquad
 \E Z_{\mathrm{miss}}\le4(m\log2+1).
\]
\end{lemma}

\begin{proof}
We choose an exponential parameter separately for small-set excess,
whole-set excess, and omitted-set deficit, and then take the
required maxima.
For a fixed ordinary set of size $s\ge2$, put
$R=m/(s-1)$ and $\theta=s\log R$.  Its internal indicator $I$ has
probability $s/(CR)$, and
\begin{equation}\label{eq:ordinary-exponential}
 \E e^{\theta(I/s-1/C)}
 \le\exp\left\{-\frac sC(\log R-1+1/R)\right\}\le1.
\end{equation}
The last inequality uses $\log R\ge1-1/R$ for $R\ge1$.
Stopping the resulting exponential supermartingale at its first
crossing, first at a bounded horizon and then taking a limit, gives
an all-time tail of $e^{-\theta x}$ for this set.  For $s\le m/2$,
$R\ge m/s\ge2$.  Since $\binom ms\le(em/s)^s$, for $x\ge3$,
\[
\begin{aligned}
 \Pr(Z_{\mathrm{small}}>x)
 &\le\sum_{s=2}^{\lfloor m/2\rfloor}
       \binom ms R^{-sx}\\
 &\le\sum_{s\ge2}(e/4)^s2^{-s(x-3)}
 \le B_0\,2^{-2(x-3)},
 \qquad B_0=\frac{(e/4)^2}{1-e/4}<2.
\end{aligned}
\]
Tail integration gives
\begin{equation}\label{eq:hub-small-tail}
 \E Z_{\mathrm{small}}\le3+\frac{B_0}{2\log2}<5.
\end{equation}

For the whole ordinary set, $I$ is Bernoulli with parameter
$(m-1)/C$.  At exponential parameter one,
\[
 \E e^{I/m-1/C}
 \le\exp\left\{\frac{-1+(m-1)(e^{1/m}-1)}C\right\}\le1,
\]
because $e^{1/m}\le m/(m-1)$.  Hence $\E F\le1$.
For $r=|B|\le m/2$, the probability of an internal request touching
$B$ is
\[
 p_B=\frac{r(2m-r-1)}{mC}\ge\frac{5r}{4C}.
\]
If $I$ is its indicator, then
\[
 \E e^{(r/C-I)/4}
 \le\exp\{r/(4C)-p_B(1-e^{-1/4})\}\le1.
\]
Here $1-e^{-1/4}\ge1/5$.  There are at most $2^m$ omitted sets, so a
union bound and tail integration give
\[
 \E Z_{\mathrm{miss}}\le4(m\log2+1).
\]
\end{proof}

\begin{lemma}[Joint all-cut workload bound]\label{lem:hub-all-cuts}
For the independently sampled near-critical family with $m\ge16$,
we have $\W\le20$.
\end{lemma}

\begin{proof}
Nearly full ordinary sets share most of their internal requests; we
retain that common part before taking a maximum over their complements.
For
$A=[m]\setminus B$ with $r=|B|<m/2$, use the exact identity
\[
 \frac{N_t(A)}{m-r}-\frac tC
 =\frac m{m-r}\left(\frac{N_t}{m}-\frac tC\right)
   +\frac{rt/C-M_t(B)}{m-r}.
\]
The empty omitted set gives just the process defining $F$.
Singleton ordinary sets have no internal requests.  Taking positive
parts and suprema therefore yields
\begin{equation}\label{eq:hub-joint-decomposition}
 Z\le Z_{\mathrm{small}}+2F+2Z_{\mathrm{miss}}/m.
\end{equation}

Lemma~\ref{lem:hub-internal} and
\eqref{eq:hub-joint-decomposition} give
\begin{equation}\label{eq:hub-ordinary-mean}
 \E Z\le7+8\log2+8/m.
\end{equation}

Let $A_i(t)$ count requests of the hub pair $\{0,i\}$, and set
\[
 Y=\max_{i\in[m]}\sup_t[t/C-A_i(t)]_+.
\]
With $\theta_*=(Cp_*)/2=(m+1/m)/2$ and $I$ Bernoulli of parameter
$p_*$,
\[
 \E e^{\theta_*(1/C-I)}
 \le\exp\{p_*/2-p_*(1-e^{-\theta_*})\}\le1.
\]
A union bound over the $m$ vertices and tail integration yield
\[
 \E Y\le\frac{2(\log m+1)}{m+1/m}\le2.
\]
For an ordinary set $A$ of size $s$,
\begin{equation}\label{eq:ordinary-cut-comparison}
 R_t(A)=\frac C{C-s}[N_t(A)/s-t/C]_+
 \le(1+1/m)Z.
\end{equation}
For a hub-containing cut $V\setminus B$, the number of requests
touching $B$ is at least $\sum_{i\in B}A_i(t)$.  The complementary
cut formula thus gives
\begin{equation}\label{eq:hub-cut-comparison}
 R_t(V\setminus B)
 \le\frac C{C-|B|}
       \left[\frac1{|B|}\sum_{i\in B}(t/C-A_i(t))\right]_+
 \le(1+1/m)Y.
\end{equation}
Every proper cut is in one of these two classes.  Because $c_{\min}=1$,
\[
 \W\le1+(1+1/m)(\E Z+\E Y)
 \le1+\frac{17}{16}\left(9+8\log2+\frac12\right)<20.
\]
\end{proof}

\begin{theorem}[Heterogeneous hub instance]\label{thm:hub}
For the independently sampled near-critical family defined above
with $m\ge16$ and empty initial loads,
\begin{equation}\label{eq:hub-workload}
 \W\le20.
\end{equation}
Every irrevocable integral trajectory satisfies
\begin{equation}\label{eq:hub-integral-lower}
 \sup_t\E H_t\ge\frac{\log m}{1024}\qquad(m\ge16).
\end{equation}
Every irrevocable fractional trajectory satisfies
\begin{equation}\label{eq:hub-fractional-lower}
 \sup_t\E H_t\ge\frac{\log m}{2048}\qquad(m\ge256).
\end{equation}
Both lower bounds allow complete advance knowledge of the requests.
In contrast, the independently reoptimized fractional prefix benchmark
satisfies
\begin{equation}\label{eq:hub-prefix}
 \E\sup_{t\ge0}\operatorname{OFF}_{\mathrm{frac}}
       (e_1,\ldots,e_t;c)\le40.
\end{equation}
\end{theorem}

\begin{proof}
The missed vertices supplied by Lemma~\ref{lem:hub-coupons} require more
external repair than the allowance $\Delta_S=1/C$ permits.
Use that lemma's $T,h,\mu$ and the set $S=[m]$.
Take $z=\mu/2$.  Since $\log m/m^{3/4}$ decreases for $m\ge16$,
\[
 \frac{T\Delta_S}{z}=\frac{2h}{\mu}
 \le\frac{\log m}{8m^{3/4}}\le\frac1{16}.
\]
Lemma~\ref{lem:budget} and \eqref{eq:hub-coupon-moments} now give
\[
 \sup_t\E H_t\ge\frac h2\left(\frac18-\frac1{16}\right)
 =\frac h{32}\ge\frac{\log m}{1024}.
\]
For $m\ge256$, $z\ge32$, so
$2C\Delta_S/(c_*(S)z)=2/z\le1/16$.  Applying
Lemma~\ref{lem:fractional-budget} instead gives
$\sup_t\E H_t\ge h/64\ge\log m/2048$.

Lemma~\ref{lem:hub-all-cuts} proves \eqref{eq:hub-workload}.
Finally, Proposition~\ref{prop:offline} gives, for every prefix,
$\operatorname{OFF}_{\mathrm{frac}}\le2\max_S R_t(S)$.
Taking the time supremum and expectation proves
\eqref{eq:hub-prefix}.
\end{proof}

The prefix benchmark in \eqref{eq:hub-prefix} permits previously
assigned jobs to be reassigned when the prefix changes.
Theorem~\ref{thm:hub} shows why that distinction matters: knowing the
future does not remove the cost of keeping earlier assignments fixed.
The example's ordinary set has relative slack $1/m$.
We next show that bounded workload also fails to control online cost
when all relative cut slacks remain uniformly positive.

\subsection{A constant-slack example}

We use a weak, explicit version of the classical layered lower-bound
argument for two choices; see Azar, Broder, Karlin, and
Upfal~\cite{ABKU99} for the foundational balanced-allocation results.
The following proof includes the arbitrary online strategies and
independent side information needed here.

\begin{lemma}[A layered lower bound]\label{lem:layered}
Let $m\ge2$ and $\ell\ge0$ be integers. Suppose $\ell$ independent
requests are uniform pairs of distinct vertices from $m$ bins. Every request
must be assigned immediately and integrally
to one of its two bins.  An online strategy may use arbitrary side information that is independent
of the entire request sequence.  For an integer $k\ge1$, put
\[
 t=\lfloor \ell/k\rfloor,\qquad
 b=\frac{t}{4km},\qquad a_i=mb^{2^i-1}.
\]
If $0<b\le1$ and $a_k\ge2$, then
\begin{equation}\label{eq:layered-general}
 \Pr\left(\max_v I_v(\ell)\ge k\right)
 \ge1-k\exp\{-tb^{2^k-2}/16\},
\end{equation}
where $I_v(\ell)$ is the number of these requests assigned to $v$.
In particular, for $\ell=\lfloor m/4\rfloor$ and
$k=\lfloor\tfrac12\log_2\log m\rfloor$, all sufficiently large $m$
satisfy
\begin{equation}\label{eq:layered-special}
 \Pr\left(\max_v I_v(\ell)\ge k\right)
 \ge1-k\exp\{-m^{3/4}/(256k)\}.
\end{equation}
\end{lemma}

\begin{proof}
Freezing the set of already heavy bins at the start of each phase makes
future requests inside that set independent trials, and every such
request must increase the load of a heavy bin.  Use the first $kt$
requests in $k$ phases of length $t$.  At the start of phase
$i\in\{0,\ldots,k-1\}$, let $S_i$ be the bins whose counted load is
at least $i$.  Suppose the past has $|S_i|\ge a_i$ and no load has yet
reached $k$.

Freeze $S_i$ for this phase.  Conditional on the past and the
independent side information, the number $B_i$ of phase requests
wholly inside $S_i$ is binomial.  For $s=|S_i|\ge a_i\ge a_k\ge2$,
\[
 \Pr(e\subseteq S_i)=\frac{s(s-1)}{m(m-1)}
 \ge\frac12(s/m)^2.
\]
Thus $\E[B_i\mid S_i]\ge t(a_i/m)^2/2$.  If a binomial variable has
mean $\mu$, exponential Markov at parameter $\log2$ gives
\[
 \Pr(B\le\mu/2)
 \le\exp\{\mu((\log2)/2-1/2)\}\le e^{-\mu/8}.
\]
Consequently, except with conditional probability at most
$\exp[-t(a_i/m)^2/16]$,
\[
 B_i\ge\frac t4(a_i/m)^2.
\]
Every one of these requests is assigned to a bin already at load at
least $i$.  If no bin reaches $k$ during the phase, each recipient
receives fewer than $k$ of these jobs.  Hence at least $B_i/k$
distinct bins end the phase at load at least $i+1$, and
\[
 |S_{i+1}|\ge\frac{B_i}{k}
 \ge\frac{t}{4k}(a_i/m)^2=a_{i+1}.
\]
Initially $|S_0|=m=a_0$.  If every phase estimate succeeds, either a
bin reaches $k$ earlier or $|S_k|\ge a_k>0$ at the end.  Since
$a_i$ decreases with $i$, each conditional phase-failure probability
is at most $\exp[-tb^{2^k-2}/16]$.  A union bound proves
\eqref{eq:layered-general}.

For the specialization, eventually $t\ge m/(16k)$ and
$b\ge1/(64k^2)$.  With $\Lambda=\log m$ and the stated choice of $k$,
\[
 (2^k-1)\log(64k^2)
 =O(\sqrt\Lambda\log\log\Lambda)=o(\Lambda).
\]
Thus $a_k\ge m^{3/4}$ for all sufficiently large $m$.
For an explicit sufficient range, $\Lambda\ge4096$ implies
$64k^2\le\Lambda$ and $\sqrt\Lambda\log\Lambda\le\Lambda/4$.
Since $b\le1$,
\[
 \frac t{16}b^{2^k-2}
 =\frac t{16}\frac{a_k}{mb}
 \ge\frac{m^{3/4}}{256k}.
\]
This proves \eqref{eq:layered-special}.
\end{proof}

For the constant-slack family, again give each ordinary vertex capacity
one and the hub capacity $m^2$, with $C=m^2+m$.  With probability
$\varepsilon=m/(2C)$, request a uniform distinct ordinary pair; otherwise
request a uniform hub pair.  Here $m\ge2$.

\begin{lemma}[Constant slack and bounded workload]\label{lem:slack-workload}
For the independently sampled constant-slack family with $m\ge2$,
\begin{equation}\label{eq:constant-slack-workload}
 \W\le10,
 \qquad
 \min_{\varnothing\ne S\subsetneq V}
 \frac{c(S)/C-\Pr(e\subseteq S)}
      {\min\{c(S)/C,c(S^c)/C\}}=\frac12.
\end{equation}
\end{lemma}

\begin{proof}
The internal stream uses only half the ordinary set's capacity
rate, allowing the same exponential estimate to control ordinary sets
of every size directly.
For an ordinary set of size $s$,
\[
 q_s=\frac{s(s-1)}{2C(m-1)},\qquad
 \frac{s/C-q_s}{s/C}=1-\frac{s-1}{2(m-1)}\ge\frac12.
\]
The ordinary side has capacity share at most $m/C\le1/2$.
A particular hub pair has probability
\[
 p_*=(1-\varepsilon)/m=\frac{m^2+m/2}{Cm}.
\]
For an ordinary set $B$ of size $s$, the probability that a request
touches $B$ is
\[
 p_{\mathrm{hit}}(B)=sp_*+
 \varepsilon\frac{s(2m-s-1)}{m(m-1)}.
\]
Consequently the slack of $V\setminus B$, divided by the smaller side's
capacity share $s/C$, is
\[
 \frac{p_{\mathrm{hit}}(B)-s/C}{s/C}
 =m+\frac{m-s}{2(m-1)}\ge1.
\]
The ordinary set $[m]$ attains relative slack $1/2$.
This proves the relative-slack assertion.

Set
\[
 Z=\sup_t\max_{A\subseteq[m],\,|A|\ge2}
       [N_t(A)/|A|-t/C]_+.
\]
For $s=|A|\ge2$, put
$R_s=2(m-1)/(s-1)$ and $\theta_s=s\log R_s$.
The internal indicator has probability $q_s=s/(CR_s)$.
The calculation in \eqref{eq:ordinary-exponential}, with $R=R_s$,
therefore gives the all-time crossing bound $e^{-\theta_sx}$.  As $R_s\ge2m/s$, for $x\ge2$,
\[
\begin{aligned}
 \Pr(Z>x)
 &\le\sum_{s=2}^m\binom ms(2m/s)^{-sx}\\
 &\le\sum_{s\ge2}(e/4)^s2^{-s(x-2)}
 \le B_0\,2^{-2(x-2)},
 \qquad B_0=\frac{(e/4)^2}{1-e/4}<2.
\end{aligned}
\]
Thus $\E Z\le2+B_0/(2\log2)<4$.

Let $A_i(t)$ count the hub pair $\{0,i\}$, and put
\[
 Y=\max_{i\in[m]}\sup_t[t/C-A_i(t)]_+.
\]
At $\theta_*=(Cp_*)/2=(m+1/2)/2>\log2$,
\[
 \E e^{\theta_*(1/C-I)}
 \le\exp\{p_*/2-p_*(1-e^{-\theta_*})\}\le1.
\]
The union bound and tail integration give
\[
 \E Y\le\frac{2(\log m+1)}{m+1/2}\le2.
\]
The cut comparisons \eqref{eq:ordinary-cut-comparison} and
\eqref{eq:hub-cut-comparison} depend only on the capacities and on
hub offers being a subset of all offers to the omitted set.  They
therefore remain valid for this law.  Ordinary singletons contribute no
positive workload.
It follows that
\[
 \W\le1+(1+1/m)(\E Z+\E Y)
 \le1+\frac32(4+2)=10.
\]
\end{proof}

\begin{theorem}[A constant-slack online obstruction]\label{thm:slack}
For the independently sampled constant-slack family, started empty,
$\W\le10$ and the relative cut slack is exactly $1/2$. For all
sufficiently large integers $m$, every online integral strategy
satisfies, at the deterministic time $C=m^2+m$,
\begin{equation}\label{eq:constant-slack-lower}
 \Pr\{H_C\ge k-1\}
 \ge1-e^{-m/16}-k\exp\{-m^{3/4}/(256k)\},
 \qquad k=\left\lfloor\tfrac12\log_2\log m\right\rfloor.
\end{equation}
In particular, $\OPT=\Omega(\log\log m)$.
\end{theorem}

\begin{proof}
The normalized mean is one at time $C$, whereas the internally forced
subsequence already creates a large maximum.
The number of internal requests by that time is binomial with mean
$C\varepsilon=m/2$.  Therefore
\[
 \Pr\{N_C([m])<\lfloor m/4\rfloor\}\le e^{-m/16}.
\]
Generate the sequence by first choosing all internal/hub types and
then the conditional request labels.  Conditional on the entire type
schedule and all hub labels, the internal labels remain independent
uniform ordinary pairs.  These conditioned variables, and the strategy's
independent random seed, are independent side information for the
internal allocation problem.  On the event
$N_C([m])\ge\lfloor m/4\rfloor$, the decisions on the first
$\lfloor m/4\rfloor$ internal requests form an online strategy covered by
Lemma~\ref{lem:layered}.  Other assignments can only increase the
ordinary loads.  Outside the two exceptional events in
\eqref{eq:constant-slack-lower}, some ordinary vertex therefore has
load at least $k$.  The capacity-weighted normalized mean at time $C$
is one, so the minimum normalized load is at most one.  This proves
\eqref{eq:constant-slack-lower}.

Lemma~\ref{lem:slack-workload} supplies the workload and slack assertions.
Finally, the right side of \eqref{eq:constant-slack-lower} tends to one,
while $k=\Theta(\log\log m)$.  Taking expectations proves the stated
lower bound on $\OPT$.
\end{proof}

\begin{remark}[Scope of the examples]
Both constructions use unequal capacities that are not proportional to
vertex offer probabilities.  They therefore leave open whether the
stochastic cut workload controls the optimal gap up to a universal
constant for equal or offer-proportional capacities.  The examples also
separate two issues: constant cut slack does not prevent an online
allocation cost, and advance knowledge does not remove the cost of
irrevocability in the near-critical construction.
\end{remark}

\section{Open problems}

The cycle bounds leave the question whether
$\OPT(C_n)=\Theta(\log n)$.  Corollary~\ref{cor:path-cycle} shows that
an $O(\log n)$ cycle bound is equivalent to a uniform logarithmic bound
for all trees with degree-proportional capacities.  Corollary~\ref{cor:tree-workload}
identifies this as the workload scale on trees.

A broader question is whether a universal constant $K$ satisfies
$\OPT\le K\W$ for every strictly feasible equal-capacity instance.
The same question can be asked for connected pair-request laws with
offer-proportional capacities $c_v=(n/2)\Pr(v\in e_1)$.
The examples in Section~\ref{sec:heterogeneous} do not belong to either
class.  By Corollary~\ref{cor:batch}, these questions are equivalently
constant-factor comparisons with the expected largest separately
optimized prefix gap.

For regular graphs, is an upper bound
$O((d/\kappa)\log n)$ always attainable?
Theorem~\ref{thm:rings} supplies families requiring this product scale,
while Theorem~\ref{thm:regular} leaves the routing factor $\alpha_G$.

\end{document}